\documentclass[apj]{emulateapj}

\usepackage{amssymb,amsmath}      
\usepackage{natbib}
\usepackage[breaklinks,colorlinks=true,linkcolor=red,citecolor=blue]{hyperref}      

\usepackage{xcolor}
\usepackage{soul}
\usepackage{multirow}

\newcommand{\question}[1]{~{\bf ???}~}

\shorttitle{Estimating lifetimes of UV-selected massive galaxies}
\shortauthors{Lin et al.}

\begin{document}

\title{Estimating lifetimes of UV-selected massive galaxies at $0.5\le z\le2.5$ in the COSMOS/UltraVISTA field through clustering analyses}

\author{Xiaozhi Lin\altaffilmark{1,2}, Guanwen Fang\altaffilmark{3,4}, Yongquan Xue\altaffilmark{1,2}, Lulu Fan\altaffilmark{1,2}, Xu Kong\altaffilmark{1,2}}

\altaffiltext{1}{CAS Key Laboratory for Research in Galaxies and Cosmology, Department of Astronomy, University of Science and Technology of China, Hefei 230026, China; xuey@ustc.edu.cn}
\altaffiltext{2}{School of Astronomy and Space Sciences, University of Science and Technology of China, Hefei 230026, China}
\altaffiltext{3}{School of Mathematics and Physics, Anqing Normal University,  Anqing 246133, China; wen@mail.ustc.edu.cn}
\altaffiltext{4}{Guanwen Fang and Xiaozhi Lin contributed equally to this work.}
\email{xuey@ustc.edu.cn, wen@mail.ustc.edu.cn}

\begin{abstract}
We present an estimation of lifetimes of massive galaxies with distinct UV colors at $0.5\le z\le2.5$ in the COSMOS/UltraVISTA field. After dividing the galaxy sample into subsamples of red sequence (RS), blue cloud (BC), and green valley (GV) galaxies in different redshift bins, according to their rest-frame extinction-corrected UV colors, we derive their lifetimes using clustering analyses. Several essentials that may influence the lifetime estimation have been explored, including the dark matter (DM) halo mass function (HMF), the width of redshift bin, the growth of DM halos within each redshift bin, and the stellar mass. We find that the HMF difference results in scatters of $\rm{\sim0.2~dex}$ on lifetime estimation; adopting a redshift bin width of $\Delta z=0.5$ is good enough to estimate the lifetime; and no significant effect on lifetime estimation is found due to the growth of DM halos within each redshift bin. The galaxy subsamples with higher stellar masses generally have shorter lifetimes; however, the lifetimes among different subsamples at $z>1.5$ tend to be independent of stellar mass. Consistently, the clustering-based lifetime for each galaxy subsample agrees well with that inferred using the spectral energy distribution modeling. Moreover, the lifetimes of the RS and BC galaxies also coincide well with their typical gas depletion timescales attributed to the consumption of star formation. Interestingly, the distinct lifetime behaviors of the GV galaxies at $z\le1.5$ and $z>1.5$ can not be fully accounted for by their gas depletion timescales. Instead, this discrepancy between the lifetimes and gas depletion timescales of the GV galaxies suggests that there are additional physical processes, such as feedback of active galactic nuclei, accelerating the quenching of GV galaxies at high redshifts.
\end{abstract}


\section{Introduction}
\label{sect:intro}

Understanding complex physical processes involved in star formation of galaxies is one of the main goals in modern astronomy.
On this subject, one of the key questions is how long star-forming galaxies (SFGs) persist in forming stars, to which clues have been provided by studying the evolution of stellar mass functions (SMFs) of SFGs and quiescent galaxies (QGs).
In the past decades, thanks to the large-scale spectroscopic and photometric surveys, such as the Sloan Digital Sky Survey (SDSS), NMBS, zCOSMOS, UltraVISTA, and zFOURGE, robust SMFs and luminosity functions of both SFGs and QGs have been established, which give us an understanding of the evolution of their space densities \citep[e.g.,][]{Moustakas_2013,Muzzin_2013b,Tomczak_2014}. Recent extensive surveys, e.g., BOSS that covers a total of over $\rm{10,000~deg^2}$ \citep{Eisenstein_2011,Dawson_2013}, enable us to study the SMFs of SFGs and QGs with stellar masses of $M_*>10^{11.5}~M_\odot$ and at $z<0.7$. Future surveys, such as the ongoing eBOSS \citep{Dawson_2016,Blanton_2017} and DESI \citep{Desi_2016}, would have the potential to study the SMFs of SFGs and QGs at higher redshifts.

Among these findings, it has usually been found that changes of star-formation rates (SFRs) of galaxies are accompanied by that of their colors \citep[e.g.,][]{Baldry_2004,Bell_2004,Faber_2007}. According to a bimodality in the color-magnitude diagram and thereby defining three distinct galaxy populations, i.e., red sequence (RS), blue cloud (BC), and green valley (GV) galaxies, astronomers have systematically studied multiple physical properties of these three galaxy populations. Though some dust-reddened SFGs may contaminate the red population, most RS galaxies generally exhibit low levels of star formation and old stellar populations, while the BC population is usually dominated by those galaxies with high SFRs and young galaxies with plenty of gas. The nature of GV galaxies lies between the other two galaxy populations \citep[e.g.,][]{Gu_2018,Gu_2019}.

Apart from the galaxy colors \citep[e.g.,][]{Bell_2004,Baldry_2004,Borch_2006,Xue_2010,Salim_2014,Lee_2015,Wang_2017}, there are other indicators characterizing the star-formation nature of galaxies, e.g., the morphology \citep{Strateva_2001,Pan_2013,Barro_2013,Barro_2014} and some spectral features such as the Balmer absorption line \citep[e.g.,][]{Kuntschner_2002,Kim_2013,Kim_2017} and the $4000~\AA$ break \citep[e.g.,][]{Kauffmann_2003,Lambas_2012,Rowlands_2018,Angthopo_2019}. Using these indicators, a number of studies have found that the color or SFR transition of galaxies is associated with the consumption of their gas content \citep[e.g.,][]{Keres_2005,Dekel_2006,Kruijssen_2015,Nelson_2018}. If this ``quenching'' scenario is correct, the timescales of the galaxy color/SFR transition and their gas depletion must be consistent.
Some authors have studied galaxy quenching by quantifying the evolution of SMFs (or luminosity functions) of SFGs and QGs at high redshifts \citep[e.g.,][]{Fritz_2014,Rowlands_2018}. The others, with the help of some infrared/sub-millimeter surveys (e.g., Hershel and ALMA), have managed to study the cosmic evolution of the gas content as well as the gas depletion rate of galaxies with different SFRs \citep[e.g.,][]{Lagos_2011,Geach_2011,Tacconi_2018,Liu_2019,Castignani_2020,Magnelli_2020}. These studies have revealed that both the galaxy quenching and gas depletion timescales increase with decreasing redshifts. The typical quenching timescale of GV galaxies can be as long as several Gyrs in the local universe \citep{Rowlands_2018,Phillipps_2019,Correa_2019}. Methods such as fitting the spectral energy distributions (SEDs) of galaxies \citep[e.g.,][]{Phillipps_2019,Zick_2018,Belfiore_2018} and using cosmological simulations \citep[e.g.,][]{Feldmann_2017,Nelson_2018,Correa_2019,Donnari_2019} have been widely used to derive the star-formation histories of galaxies and their lifetimes.

Although great efforts have been made, there are still some challenges to understand the distinct star-formation histories of different galaxy populations. For example, it is difficult to obtain high-quality spectra for galaxies with wide ranges of SFR and redshift. While there have been many robust optical spectroscopic surveys in the local universe \citep[e.g.,][]{Abolfathi_2018,Aguado_2019a,Aguado_2019b,Westfall_2019}, obtaining high-resolution spectra is still a challenging task when going to higher redshifts, where the galaxy sample may suffer from selection effects to a large degree \citep[e.g.,][]{Gruppioni_2019,Zheng_2020}. Besides, using photometric colors to separate SFGs from the quiescent population may also suffer from contamination of dust-reddened galaxies. Moreover, it is difficult to systematically measure gas properties, e.g., through the 21~cm and carbon monoxide molecule (CO) emission lines, for a large number of high-redshift galaxies over a wide range of SFR. To estimate the gas content of high-redshift galaxies, researchers usually rely on several assumptions, including, e.g., a constant CO-to-$\rm{H_{2}}$ conversion factor \citep{Bolatto_2013,Tacconi_2013,Shangguan_2020}, a constant gas-to-dust mass ratio as found in the local universe \citep[e.g.,][]{Leroy_2011}, and multiple dust natures (e.g., temperature and optical properties) \citep{Scoville_2016,Scoville_2017}.

Considering the aforementioned challenges and uncertainties, here we choose to take another route, i.e., using clustering analyses, to calculate the lifetimes of galaxies with different levels of star formation and at a broad range of redshift (0.5$\le z \le$2.5).
\citet{Hickox_2012} used their clustering results to estimate the lifetimes of sub-millimeter galaxies based on the assumption that for each DM halo with a mass similar to that of the sub-millimeter galaxies, the galaxy reside in it would undergo the sub-millimeter phase.
This assumption is based on the scenario of co-evolution of galaxy/AGN and DM halo \citep[e.g.,][]{Hickox_2009} that the behaviors of AGN host galaxies are accompanied by the growth of their host dark matter (DM) halos. When the halo mass reaches a certain critical value, the star-formation/AGN feedback will be triggered, and thus the state of the AGN host galaxy changes (see their Figure 16). This scenario is supported by a number of studies, where the DM halo plays an important role in shaping the star-formation activity and regulating the central black hole activity \citep[e.g.,][]{Moster_2013,Cai_2013,Behroozi_2013,Vogelsberger_2014,Chen_2019}.
This method has also been applied to estimate the duty cycle of high-redshift AGNs \citep[e.g.,][]{Martini_2001,Haiman_2001,Eftekharzadeh_2015,He_2017}.

In this work, we make a follow-up estimation of the lifetimes of BC, GV, and RS galaxies based on the clustering results of the three galaxy populations in \citet{Lin_2019}.
We inherit the classification of the three galaxy populations in \citet{Wang_2017}, where they demonstrated that their extinction-corrected UV-color separation of the BC, GV, and RS galaxies can be a good representation of the intrinsic star-forming, transition, and quiescent galaxies.
Apart from \citet{Lin_2019}, the higher clustering (larger halo masses) of quiescent galaxies than star-forming galaxies had also been uncovered
by a number of studies\citep[e.g.,][]{Norberg_2002,Coil_2004,Coil_2008,Zehavi_2011,Bray_2015,Coil_2017}, thus we would expect a clear difference in their estimated lifetimes according to the galaxy-halo co-evolution scenario \citep[e.g.][]{Hickox_2009} and the assumption of \citet{Hickox_2012}.


The method we apply for the lifetimes of UV-selected galaxies is similar to that of \citet{Hickox_2012} for sub-millimeter galaxies. But in \citet{Hickox_2012}, they only used a single halo mass function (HMF) to estimate the clustering-based lifetime of galaxies. In this work, we further check and evaluate to which extent different factors, such as the choices of the HMF, the redshift bin width, the growth of DM halos within each redshift bin, and the stellar mass of galaxy sample would affect the lifetimes of different galaxy populations.
Besides, it is meaningful to compare our lifetime result with that derived from SED modeling, since clustering measurement and SED modeling are two entirely different routes.
Moreover, recent mid-to-far infrared observations have enabled us to make rough estimation of the molecular gas amounts and gas depletion timescales through the consumption of star formation in different types of galaxies \citep[e.g.,][]{Popping_2017,Tacconi_2018,Catinella_2018,Fluetsch_2019,Decarli_2019,Magnelli_2020}.
Thus, comparing the lifetimes with the typical gas depletion timescales in these galaxies would shed light on the physical connections between galaxy color transformation and gas consumption mechanisms as well.

The outline of this paper is the following. We briefly describe our adopted data and sample selection in Section~\ref{sect:data}. In Section~\ref{sect:method}, we introduce the clustering method to calculate the lifetimes of the UV-selected galaxies based on their clustering properties presented in \citet{Lin_2019}. Then we investigate several factors affecting the lifetime estimation in Section~\ref{sect:Dis}. We further compare our clustering-based lifetimes with those derived from the SED modeling as well as the typical gas depletion timescales of galaxies in Section~\ref{sect:modeling}. Finally, Section~\ref{sect:sum} summarizes our major results.

Throughout this paper, we adopt a flat cosmology with $\Omega_{\rm M}=0.3$, $\Omega_{\Lambda}=0.7$, $H_0=70~{\rm km~s^{-1}~Mpc^{-1}}$, and a normalization of $\sigma_8=0.84$ for the matter power spectrum.

\section{Data}\label{sect:data}

The data analyzed in this work are from the COSMOS/UltraVISTA survey, which covers $\rm{\sim1.62~deg^2}$ in the COSMOS/UltraVISTA field. We utilize the \citet{Muzzin_2013a} catalog, which is based on the UltraVISTA $K_s$-band imaging \citep{McCracken_2012} and includes multi-wavelength photometries, i.e., 30 bands from $\sim0.15$ to $\sim24~\mu$m. The photometric redshift, rest-frame colors, and stellar population parameters, including the stellar mass and star-formation history, are estimated for each galaxy by the SED-fitting method \citep{Muzzin_2013a}.

\subsection{Redshift, Stellar Mass, and $A_{V}$}

In the following, we directly adopt the photometric redshift, stellar mass, star-formation history, attenuation, and rest-frame colors estimated by \citet{Muzzin_2013a}. Only relevant data are briefly introduced as follows (see \citealt{Muzzin_2013a} for more details).

The \citet{Muzzin_2013a} photometric catalog contains two bands from GALEX (FUV and NUV), one band from CFHT/MegaCam ($u^{*}$), six broad bands ($g^{+}$, $r^{+}$, $i^{+}$, $z^{+}$, ${\rm B}_{j}$, and ${\rm V}_{j}$) and twelve optical medium bands (IA427--IA827) from Subaru/Suprime-Cam, four near-infrared broad bands ($Y$,  $J$, $H$, and $K_{\rm s}$) as well as the 3.6, 4.5, 5.8, 8.0, and 24 $\mu$m channels from {\it Spitzer}'s IRAC and MIPS. The observed-frame wavelength coverage of this catalog is from $\sim 0.15~\mu $m to $\sim 24~\mu$m, and then for the maximal redshift of $\sim 2.5$ considered in this paper, the rest-frame wavelength coverage is $\sim 0.04~\mu$m to $\sim 6.8~\mu$m. Therefore, the UltraVISTA data have a good wavelength coverage in the redshift desert at $z \sim 1.4-2.5$.

The photometric redshifts and rest-frame colors in the COSMOS/UltraVISTA catalog were computed through fitting the multiwavelength SEDs using EAZY software \citep{EAZY}. The templates adopted by \citet{Muzzin_2013a} include seven initial templates in EAZY (i.e., six from the PEGASE models of \citealt*{FR_1999} and one red template from the model of \citealt*{M05}) and two additional templates, i.e., a one-gigayear-old single-burst \citet{BC03} model for strong post-starburst-like features at $z>1$, and a slightly dust-reddened young population for the UV bright galaxies at $1.5<z<3.5$ to improve the accuracy of photometric redshift.
All the above templates were considered in the derivation of photometric redshifts of the COSMOS/UltraVISTA galaxies. During the computation of the photometric redshifts, they ran the EAZY software by fixing the photometric redshifts to the best observed spectroscopic redshifts to determine the zero point offsets in each photometric band. Then the photometry in each band was adjusted by the corresponding offset and EAZY was re-run, until the median offset in every filter converges.

At $z < 1.5$, the photometric redshifts are well consistent with the spectroscopic redshifts from zCOSMOS \citep{Lilly_2007,Lilly_2009} and other spectroscopic surveys \citep{Onodera_2012,Bezanson_2013,Sande_2013}, with a $3\sigma$-outlier fraction of $1.56\%$ and a low rms scatter of $\delta z/(1+z)=0.013$. At $z > 1.5$, the photometric redshifts agree well with that from the NOAO Extremely Wide-Field Infrared Imager (NEWFIRM) Medium-Band Survey \citep[NMBS,][]{Whitaker_2011}, with the fraction of the UltraVISTA sources considered to be $5\sigma$ outliers of the NMBS survey being $2.0\%$ and the rms dispersion of UltraVISTA photometric redshifts being $\delta z/(1+z)=0.026$. We note that, when available, the \citet{Muzzin_2013a} catalog adopted the spectroscopic redshifts rather than the photometric ones.

Although \citet{Muzzin_2013a} provided a $\chi^2$ value to evaluate each SED-fitting result in their catalog, we do not apply any cut on this value, since nearly $\sim 90\%$ of the sources have a reduced $\chi^2 < 2$.

We have removed stars and quasars using the flags in the catalog and further limit the redshifts of galaxies to be within $0.5\le z\le 2.5$ in our study. We obtain 126,222 sources in total, which are further divided into four smaller redshift bins, i.e., $0.5\le z<1.0$, $1.0\le z <1.5$, $1.5\le z <2.0$, and $2.0 \le z \le2.5$, in order to study the lifetime evolution of different galaxy populations.

\citet{Muzzin_2013a} had also provided the information of stellar population of COSMOS/UltraVISTA galaxies in their catalog. The stellar population parameters, including the stellar mass ($M_{\ast}$), visual attenuation ($A_{V}$), and the stellar age of galaxies, were computed using the FAST code. They directly adopted the redshifts they derived above as calibrations and inputs. Then the observed SEDs of COSMOS/UltraVISTA galaxies were fitted with the assembly of stellar population templates in the \citet{M05} model, assuming a solar metallicity, a Chabrier initial mass function \citep[][]{Chabrier_2003}, and a \citet{Calzetti_2000} dust extinction law. To construct the SED templates, \citet{Muzzin_2013a} assumed exponentially-declining star-formation histories for all COSMOS/UltraVISTA galaxies with the form of ${{\rm SFR}\propto \exp(-t/\tau)}$, where ${t}$ is the time since the onset of star formation and $\tau$ is the e-folding star-formation timescale that varies between $10^{7}\hspace{0.5mm}{\rm yr}$ and $10^{10}\hspace{0.5mm}{\rm yr}$, respectively.

\subsection{Sample Construction}\label{sect:sample_construction}

\begin{table*} \caption{Best-fit clustering properties of galaxy subsamples in the COSMOS/UltraVISTA field\label{tbl-1}}
\centering

\begin{tabular}{lcccccccccc}
\hline
\hline
Galaxy subsample &
\(N_{\rm source}^{\rm a}\) &
$\bar{z}^{\rm b}$ &
$\log(M_\ast/M_\odot)^{\rm c}$ &
$A^{\rm d}_\omega$ &
\(b_{\rm gal}^{\rm e}\) &
$\log(M_{\rm halo}/[h^{-1}{M_\odot}])^{\rm f}$ \\
\hline
$0.5 \le z < 1.0$ & \multicolumn{6}{l}{ $\log(M_\ast/M_\odot) \ge 10$ \& 80\% mass-completeness } &  \\
$\rm RS$ & 6051 & 0.823 & 10.53 & 0.204$\pm$0.034 & 1.87$\pm$0.16 & \(12.98_{-0.16}^{+0.14}\)   \\
$\rm GV$ & 4116 & 0.789 & 10.39 & 0.107$\pm$0.013 & 1.38$\pm$0.08 & \(12.40_{-0.15}^{+0.13}\)    \\
$\rm BC$ & 3774 & 0.843 & 10.21 & 0.092$\pm$0.011 & 1.23$\pm$0.07 & \(12.03_{-0.18}^{+0.15}\)    \\
 & \multicolumn{6}{c}{ $\log(M_\ast/M_\odot) \ge 10.5$ \& 98\% mass-completeness } &  \\
$\rm RS$ & 3234 & 0.819 & 10.80 & 0.201$\pm$0.034 & 1.80$\pm$0.15 & \(12.92_{-0.17}^{+0.14}\)   \\
$\rm GV$ & 1546 & 0.800 & 10.67 & 0.115$\pm$0.020 & 1.41$\pm$0.12 & \(12.43_{-0.22}^{+0.19}\)    \\
$\rm BC$ & 537 & 0.856 & 10.59 & 0.123$\pm$0.030 & 1.36$\pm$0.17 & \(12.28_{-0.37}^{+0.27}\)    \\
\hline
$1.0 \le z < 1.5$ & \multicolumn{6}{l}{ $\log(M_\ast/M_\odot) \ge 10$ \& 80\% mass-completeness } & \\
$\rm RS$ & 5099 & 1.223 & 10.51 & 0.149$\pm$0.031 & 2.09$\pm$0.22 & \(12.76_{-0.20}^{+0.17}\)    \\
$\rm GV$ & 2143 & 1.219 & 10.44 & 0.158$\pm$0.022 & 2.13$\pm$0.15 & \(12.80_{-0.13}^{+0.12}\)    \\
$\rm BC$ & 5798 & 1.270 & 10.26 & 0.096$\pm$0.011 & 1.67$\pm$0.10 & \(12.27_{-0.13}^{+0.12}\)    \\
 & \multicolumn{6}{c}{ $\log(M_\ast/M_\odot) \ge 10.5$ \& 98\% mass-completeness } &  \\
$\rm RS$ & 2602 & 1.222 & 10.76 & 0.188$\pm$0.043 & 2.29$\pm$0.26 & \(12.92_{-0.21}^{+0.18}\)    \\
$\rm GV$ & 921 & 1.216 & 10.69 & 0.176$\pm$0.025 & 2.20$\pm$0.16 & \(12.85_{-0.13}^{+0.12}\)    \\
$\rm BC$ & 1192 & 1.297 & 10.63 & 0.183$\pm$0.026 & 2.26$\pm$0.16 & \(12.85_{-0.13}^{+0.11}\)    \\
\hline
$1.5 \le z < 2.0$ & \multicolumn{6}{l}{ $\log(M_\ast/M_\odot) \ge 10$ \& 80\% mass-completeness } & \\
$\rm RS$ & 3354 & 1.729 & 10.50 & 0.168$\pm$0.032 & 3.04$\pm$0.29 & \(12.91_{-0.16}^{+0.14}\)    \\
$\rm GV$ & 1393 & 1.727 & 10.38 & 0.112$\pm$0.016 & 2.52$\pm$0.18 & \(12.61_{-0.13}^{+0.12}\)    \\
$\rm BC$ & 4971 & 1.733 & 10.29 & 0.119$\pm$0.014 & 2.44$\pm$0.14 & \(12.55_{-0.11}^{+0.10}\)    \\
 & \multicolumn{6}{c}{ $\log(M_\ast/M_\odot) \ge 10.5$ \& 98\% mass-completeness } &  \\
$\rm RS$ & 1719 & 1.735 & 10.74 & 0.221$\pm$0.053 & 3.27$\pm$0.39 & \(13.02_{-0.20}^{+0.16}\)    \\
$\rm GV$ & 506 & 1.735 & 10.69 & 0.151$\pm$0.037 & 2.73$\pm$0.33 & \(12.74_{-0.22}^{+0.18}\)    \\
$\rm BC$ & 1316 & 1.780 & 10.67 & 0.144$\pm$0.015 & 2.58$\pm$0.13 & \(12.64_{-0.09}^{+0.09}\)    \\
\hline
$2.0 \le z \le 2.5$ & \multicolumn{6}{l}{ $\log(M_\ast/M_\odot) \ge 10$ \& 80\% mass-completeness } &  \\
$\rm RS$ & 1251 & 2.231 & 10.54 & 0.294$\pm$0.042 & 5.50$\pm$0.39 & \(13.36_{-0.10}^{+0.09}\)   \\
$\rm GV$ & 1029 & 2.243 & 10.46 & 0.147$\pm$0.022 & 3.57$\pm$0.27 & \(12.75_{-0.12}^{+0.11}\)   \\
$\rm BC$ & 2239 & 2.215 & 10.31 & 0.211$\pm$0.025 & 3.59$\pm$0.21 & \(12.78_{-0.10}^{+0.09}\)   \\
 & \multicolumn{6}{c}{ $\log(M_\ast/M_\odot) \ge 10.5$ \& 98\% mass-completeness } &  \\
$\rm RS$ & 697 & 2.264 & 10.71 & 0.374$\pm$0.084 & 5.67$\pm$0.64 & \(13.40_{-0.15}^{+0.13}\)   \\
$\rm GV$ & 477 & 2.224 & 10.69 & 0.286$\pm$0.057 & 4.90$\pm$0.49 & \(13.21_{-0.14}^{+0.12}\)   \\
$\rm BC$ & 662 & 2.194 & 10.70 & 0.259$\pm$0.040 & 4.38$\pm$0.34 & \(13.07_{-0.11}^{+0.10}\)   \\
\hline
\end{tabular}

\begin{tabular}{p{17cm}}
$^{\rm a}$ The source number of each galaxy subsample. \\
$^{\rm b}$ The median redshift of each galaxy subsample. \\
$^{\rm c}$ The median stellar mass of each galaxy subsample. \\
$^{\rm d}$ The clustering amplitude of the angular correlation function of each galaxy subsample. \\
$^{\rm e}$ The bias of each galaxy subsample. \\
$^{\rm f}$ The halo mass approximately converted from the bias. \\
\end{tabular}

\label{tab:galaxy_bin}
\end{table*}

Out of the above 126,222 sources, we only include the 41,218 galaxies with $\log(M_{\ast}/M_{\odot}) \ge 10$ as our final sample for further analyses. This sample is the same as that of \citet{Lin_2019}, which consists of 12 subsamples of galaxies that have three distinct UV colors in four redshift bins.
The adopted stellar-mass limit corresponds to a mass completeness of 80\% near $z \sim 2.5$, while the mass-completeness limit reaches 98\% for $\log(M_{\ast}/M_{\odot}) \ge 10.5$ (see \citealt{Lin_2019} for more details).

Utilizing the color-separation criteria proposed by \citet{Wang_2017}, we separate the RS, GV, and BC galaxies according to their rest-frame extinction-corrected UV colors.
Although there are some other color-separation criteria classifying galaxies with different SFRs, a number of them are based on the observed colors and do not consider the effect of extinction correction. Thus a proportion of RS galaxies may be imposters that are actually BC galaxies with severe dust extinction. In \citet{Brammer_2009}, they showed that by using extinction-corrected $\rm U-V$ colors, dust-reddened BC galaxies can be better separated from the intrinsic RS galaxies. Later, \citet{Wang_2017} refined the old criteria by \citet{Bell_2004} and \citet{Borch_2006} such that they are more consistent from local to $z \sim 2.5$.

\begin{figure}
\centering
\includegraphics[angle=90,width=1.1\columnwidth]{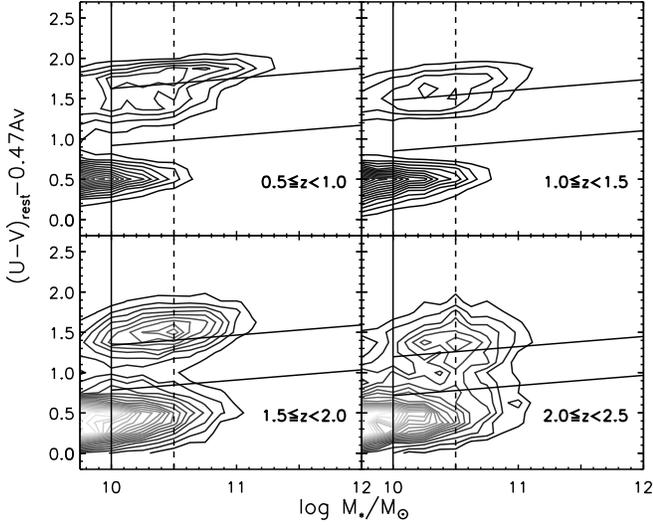}
\caption{ The distribution of COSMOS/UltraVISTA galaxies in the extinction-corrected rest-frame UV color versus stellar mass diagram. The separations of the RS, GV and BC galaxies are roughly shown as the solid lines. We have also selected subsamples of higher stellar masses (i.e., $\log(M_{\ast}/M_{\odot}) \ge 10.5$), shown as dash lines to study the influence of stellar mass on the resulting lifetimes. }
\label{fig:CMR}
\end{figure}

The new separation criteria are as follows:
\begin{equation}\label{ref:equ_sep_cri}
\begin{split}
{\rm\leftline{RS:}}\\
{\rm(U-V)_{rest}-0.47A_v \ge 0.126 \log(M_{\ast}/M_\odot)+0.58-0.286z,}\\
{\rm\leftline{GV:}}\\
{\rm(U-V)_{rest}-0.47A_v < 0.126 \log(M_{\ast}/M_\odot)+0.58-0.286z}\\
 \centerline\&\\
{\rm(U-V)_{rest}-0.47A_v \ge 0.126 \log(M_{\ast}/M_\odot)-0.24-0.136z,}\\
{\rm\leftline{BC:}}\\
{\rm(U-V)_{rest}-0.47A_v < 0.126 \log(M_{\ast}/M_\odot)-0.24-0.136z},
\end{split}
\end{equation}
where ${\rm(U-V)_{rest}}$ is the rest-frame UV color and the value of 0.47 is the correction factor computed for the \citet{Calzetti_2000} extinction law. \citet{Brammer_2009} showed that dust-reddened SFGs can be better separated from the intrinsic old and red galaxies by applying this correction factor.
According to the above criteria, we classify each galaxy into one of the three galaxy populations, i.e., RS, GV, or BC.
Then each final galaxy subsample is constructed such that the redshifts of its galaxies fall within one of the four redshift intervals between $0.5\le z\le2.5$, their stellar masses satisfy $\log(M_{\ast}/M_{\odot}) \ge 10.0$ and their rest-frame extinction-corrected UV colors belong to one of the three populations. We have also selected galaxy subsamples of $\log(M_{\ast}/M_{\odot}) \ge 10.5$ to study the influence of stellar mass on the resulting lifetimes.
The source number and median redshift of each galaxy population within each redshift bin are tabulated in Table~\ref{tab:galaxy_bin}, and the distribution of the COSMOS/UltraVISTA galaxies in the color-mass diagram as well as the rough separations of the galaxy subsamples are shown in Figure~\ref{fig:CMR}.

In \citet{Lin_2019}, we have measured the clustering amplitudes of the three galaxy populations through the two-point angular correlation functions.
Then by properly comparing their correlation functions with that of the DM halos at the respective redshifts, we have derived the bias factors of all galaxy populations.
After that, the bias of each galaxy population is converted into the corresponding host halo mass, $M_{\rm halo}$, following the prescription of \citet{Sheth_2001}. The clustering amplitudes ($A_{\omega}$), bias factors as well as the DM halo masses of all galaxy populations are also tabulated in Table~\ref{tab:galaxy_bin}.
More details on the clustering analyses are elaborated in \citet{Lin_2019}.
Note that, for easy reference, Table~\ref{tab:galaxy_bin} shown in this work is a simple replication of part of the Table~1 in \citet{Lin_2019}.

\section{Method to calculate lifetime}\label{sect:method}

In this section, we introduce the clustering method to calculate the galaxy lifetime following \citet{Hickox_2012}.
Similar to the assumption in \citet{Hickox_2012} (see their Section 4.4), we assume that for every DM halo with a mass similar to that of the measured host halo mass of RS, BC or GV galaxies in \citet{Lin_2019} (tabulated in Table~\ref{tab:galaxy_bin} in this paper), the galaxy residing in that DM halo would go through the corresponding RS, BC or GV phase during its evolution. In this assumption, we have pre-assumed a one to one halo occupation relation for the host DM halo in the halo mass range of the subsamples we concern. And we define the mass of a halo similar to the measured $M_{\rm halo}$ of one of the RS, BC or GV galaxy subsamples as the mass falling within the measured $M_{\rm halo}$ error bar of that galaxy population.

We note that fixing the mean halo occupation to one may not always be true, if we consider more complex clustering properties for these galaxies.
In some halo occupation distribution (HOD) models \citep[e.g.,][]{Zheng_2005,Zheng_2007,Ross_2009,Zehavi_2011,Wake_2011,Kim_2015,BeltzMohrmann_2020}, galaxies are assumed to occupy all DM halos above some minimum halo mass.
Thus in these models, the mean occupation of DM halo is no longer equal to one, but it is a function of $M_{\rm halo}$, i.e., more massive halos have the potential to host more galaxies.
This is also consistent with some semi-analytic prescriptions based on large cosmological simulations \citep[e.g.,][]{Berlind_2002,Manera_2013}.
But in \citet{Lin_2019}, we only adopted a simple linear bias relative to dark matter to convert the clustering amplitude into $M_{\rm halo}$, and in this simplified case, all galaxies in a given subsample reside in DM halos of similar masses.
At the low mass end (e.g., $\log(M_{\rm halo}/[h^{-1}{M_\odot}])\le13.0$), a one to one occupation relation between galaxy and DM halo may slightly over-estimate the mean number of galaxies occupying a DM halo predicted by HOD models and semi-analytic prescriptions based on cosmological simulations \citep[e.g.,][]{Zheng_2005,Manera_2013,Wechsler_2018}.
While at the high mass end (e.g., $\log(M_{\rm halo}/[h^{-1}{M_\odot}])>13.0$), the situation is the contrary, the one to one occupation relation would probably under-estimate the mean halo occupation.


Based on the assumptions above, the lifetime of a given galaxy subsample is defined as the average timescale that galaxies in the similarly massive DM halos of the subsample spend in the phase of the corresponding extinction-corrected rest-frame UV color\footnote{If average more or fewer galaxies reside in the DM halos experience the phase of the UV color of the subsample, the estimated lifetime will be correspondingly shorter or longer, respectively.}, such that,
\begin{equation}\label{eq:life_time}
lifetime_{\rm sample} = \Delta t\frac{N_{\rm sample}}{N_{\rm halo}},
\end{equation}
where $\Delta t$ is the time interval corresponding to the redshift range of the galaxy sample, $N_{\rm sample}$ is the source number of the galaxy sample, and $N_{\rm halo}$ is the number of the DM halos that have a similar mass to the halo mass of the sample.
In equation~\ref{eq:life_time}, $N_{\rm halo}$ is not necessarily equivalent to $N_{\rm sample}$.
This is because although we assume that each galaxy in the similar DM halos would undergo the respective phase of the sample, galaxy color is not constant during its evolution.
When we observe a galaxy whose host halo has a similar mass to the measured halo mass of a given galaxy sample, its color may have already changed (i.e., it may have already passed the phase of that galaxy sample) or its color may have not yet reached the color of the given galaxy sample (i.e., it may have not yet evolved into that phase).
Thus the lifetime here can be a rough estimate of the time that a galaxy is alive in a respective phase of color.

To calculate $N_{\rm halo}$, we should integrate the halo mass function (HMF) over the respective redshift range of the sample,
\begin{equation}\label{eq:integral_Nhalo}
N_{\rm halo}=\int_{V(z_{\rm max})}^{V(z_{\rm min})}\int_{M_{-}}^{M_{+}}\frac{dn_{\rm halo}}{dM_{\rm h}}(M_{\rm h},z)dM_{\rm h}dV,
\end{equation}
where $V(z_{\rm max})$ and $V(z_{\rm min})$ are the comoving volumes corresponding to the maximum and minimum redshifts of the sample, $M_{+}$ and $M_{-}$ correspond to the upper and lower limits of the DM halo mass of the subsamples in Table~\ref{tab:galaxy_bin}.
The HMF, $\frac{dn_{\rm halo}}{dM_{\rm h}}(M_{\rm h},z)$, follows the functional form of
\begin{equation}
\frac{dn_{\rm halo}}{dM_{\rm h}}(M_{\rm h},z) = f(\sigma)\frac{\bar{\rho}_{m}}{M_{\rm h}}\left|\frac{d~{\rm ln}~\sigma}{dM_{\rm h}}\right|,
\end{equation}
where $\sigma$ is the cosmology-dependent mass variance of the linear density field smoothed on the scale of $R(M)=(3M/4\pi\cdot\bar{\rho}_m)^{1/3}$, $\bar{\rho}_{m}$ is the mean density of the universe, and $f(\sigma)$ represents the functional form that defines a particular HMF fit. In this paper, we should explore how the form of HMF fit influences the lifetimes of our galaxy subsamples (see Section~\ref{sect:dis_HMFs}).

Beside HMF, we also discuss multiple factors that can influence the resultant lifetimes of our galaxy subsamples, including the choices of the width of redshift bin, the growth of DM halo mass, and the stellar mass of the galaxy sample.

\section{Discussion}\label{sect:Dis}

In this section, we explore several factors that may affect lifetimes of galaxy samples.

\subsection{Influence of Halo Mass Functions}\label{sect:dis_HMFs}

\begin{figure}
\centering
\includegraphics[angle=90,width=1.1\columnwidth]{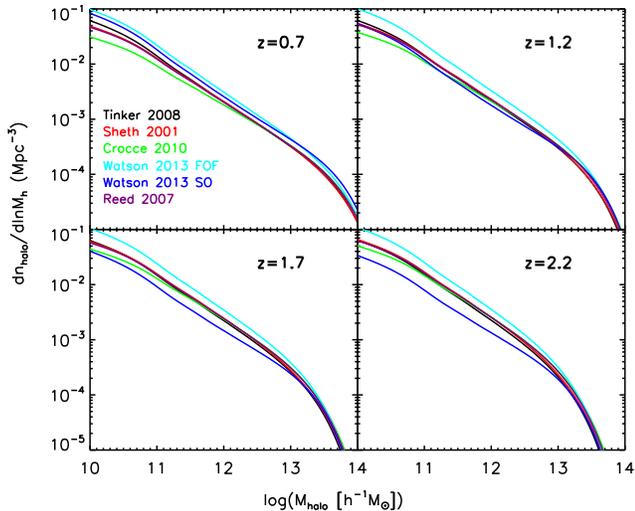}
\caption{Comparison of different HMFs at four redshifts within our concerned redshift range.}
\label{fig:HMF}
\end{figure}

The lifetimes of galaxies can be affected by different forms of HMFs to some degree.
We investigate the influence of several forms of HMFs, including that of \citet{Tinker_2008}, \citet{Sheth_2001}, \citet{Crocce_2010}, and \citet{Reed_2007}, as well as two additional forms of HMFs derived respectively by the friends-of-friends (FOF) and spherical overdensity (SO) halo-finding methods in \citet{Watson_2013}.
We list these HMFs in Table~\ref{tab:HMFs}, while a comparison of them is shown in Figure~\ref{fig:HMF}.

In general, the \citet{Watson_2013} FOF HMF shows larger numbers of halos than other HMFs at all redshifts,
so it leads to smaller lifetimes estimated through Equation~\ref{eq:life_time}.
At $z>1.5$, the \citet{Watson_2013} SO HMF appears to be the smallest among all HMFs over a large range of halo mass, but the discrepancy becomes very small towards the most massive end.

\begin{table*} \caption{Comparison of different HMFs\label{tbl-2}}
\centering

\begin{tabular}{lcccccccccc}
\hline
\hline
Ref. &
Fitting Function $f(\sigma)$ \\
\hline
\multirow{3}*{\citet{Tinker_2008}}
 & $f_{\rm T}(\sigma,z)=A((\frac{b}{\sigma})^a+1){\rm exp}\left[-\frac{c}{\sigma^2}\right],$\\
 & $A=0.186(1+z)^{-0.14}, a=1.47(1+z)^{-0.06}, b=2.57(1+z)^{-\alpha}, c=1.19$,\\
 & $\alpha={\rm exp}\left[-(\frac{0.75}{{\rm ln}(\Delta_{\rm vir}/75)})^{1.2}\right]$ \\
\hline
\citet{Sheth_2001} & $f_{\rm ST}(\sigma)=A\sqrt{\frac{2a}{\pi}}\left[1+(\frac{\sigma^2}{a\delta_c^2})^p\right]\frac{\delta_c}{\sigma}{\rm exp}\left[-\frac{a\delta_c^2}{2\sigma^2}\right]$   \\
\hline
\multirow{2}*{\citet{Crocce_2010}}
& $f_{\rm Cr}(\sigma)=A(\sigma^{-a}+b){\rm exp}\left[-\frac{c}{\sigma^2}\right],$\\
& $A=0.58(1+z)^{-0.13}, a=1.37(1+z)^{-0.15}, b=0.3(1+z)^{-0.084}, c=1.036(1+z)^{-0.024}$    \\
\hline
\multirow{2}*{\citet{Watson_2013} FOF}
& $f_{\rm W_{FOF}}(\sigma)=A((\frac{b}{\sigma})^a+1){\rm exp}\left[-\frac{c}{\sigma^2}\right],$\\
& $A=0.282, a=1.406, b=2.163, c=1.21$    \\
\hline
\multirow{6}*{\citet{Watson_2013} SO} & $f_{\rm W_{SO}}(\sigma,z)=\Gamma(\Delta,\sigma,z)A((\frac{b}{\sigma})^a+1){\rm exp}\left[-\frac{c}{\sigma^2}\right],$\\
& $(A,a,b,c)_{z=0}=(0.194,2.267,1.805,1.287),$\\
& $(A,a,b,c)_{z>6}=(0.563,874,3.810,1.453),$\\
& $(A,a,b,c)_{0<z<6}=\Omega_M(z)\times(1.907(1+z)^{-3.216}+0.074, 3.136(1+z)^{-3.058}+2.349, 5.907(1+z)^{-3.599}+2.344, 1.318),$\\
& $\Gamma(\Delta,\sigma,z)=C(\Delta)(\Delta/178)^{d(z)}{\rm exp}\left[\frac{p(1-\frac{\Delta}{178})}{\sigma^q}\right],$\\
& $C(\Delta)=0.947{\rm exp}\left[0.023(\frac{\Delta}{178}-1)\right], d(z)=-0.456\Omega_M(z)-0.139, p=0.072, q=2.130$ \\
\hline
\multirow{3}*{\citet{Reed_2007}} & $f_R(\sigma) = A\sqrt{\frac{2a}{\pi}}\left[1+\left(\frac{\sigma^2}{a\delta_c^2}\right)^p+0.6G_1+0.4G_2\right]\times\frac{\delta_c}{\sigma}{\rm exp}\left[-\frac{ca\delta_c^2}{2\sigma^2}-\frac{0.03}{(n_{\rm eff}+3)^2}(\frac{\delta_c}{\sigma})^{0.6}\right],$\\
& $A=0.310, a=0.75, ca=0.764, p=0.3, \delta_c=1.686, n_{\rm eff}=6\frac{d~{\rm ln}~\sigma^{-1}}{d~{\rm ln}~M}-3,$\\
& $G_1={\rm exp}\left[-\frac{({\rm ln}~\sigma^{-1}-0.4)^2}{0.72}\right], G_2 = {\rm exp}\left[-\frac{({\rm ln}~\sigma^{-1}-0.75)^2}{0.08}\right]$    \\
\hline
\end{tabular}

\label{tab:HMFs}
\end{table*}

Since we use a mass-selected sample, it may suffer some selection effects, e.g., the detection completeness limit of the sources, photometric uncertainties in the derivation of stellar mass using a set of stellar population models, and the effect of clustering.
In this paper, we adopt the stellar mass functions (SMFs) computed by the $1/V_{\rm max}$ method as well as their 1$\sigma$ uncertainties in the COSMOS/UltraVISTA field \citep{Muzzin_2013b} to account for the selection effects.

Here is a brief introduction of the computation of $1/V_{\rm max}$ \citet{Muzzin_2013b} SMFs in the COSMOS/UltraVISTA field.
First, they determined the maximum volume ($V_{\rm max}$) within which an object of the stellar mass of $M_\ast$ could be detected above the stellar mass completeness level.
Then they counted the galaxies in bins of $M_\ast$, and corrected those bins with $1/V_{\rm max}$.
The 1$\sigma$ uncertainties in the $1/V_{\rm max}$ \citet{Muzzin_2013b} SMFs are made up of three parts: Poisson error when counting galaxies within $V_{\rm max}$, errors from photometric uncertainties, and cosmic variance due to galaxy clustering.
To estimate the error from photometric uncertainties, they performed 100 Monte Carlo (MC) realizations of the catalog.
Within each realization, they allowed the photometry in the catalog to perturb according to the measured photometric uncertainties.
Then these 100 MC catalogs were used to recalculate SMFs, and they determined the empirical uncertainties in the SMFs due to photometric uncertainties as the range of these recalculated SMFs.
For the estimation of cosmic variance, they adopted the prescriptions of \citet{Moster_2011}.

The 1$\sigma$ uncertainties of the \citet{Muzzin_2013b} SMFs can give us a rough estimation of selection effects, reflected by fluctuations in the number counts of all galaxy subsamples.
We integrate the upper and lower limits of SMFs to obtain the 1$\sigma$ fluctuations of the galaxy amounts in each redshift bin.
Then we assume the over/under-estimated amount of sources to share the same distributions of AGN fraction and extinction-corrected UV color as that of normal galaxies in that redshift bin.
Thus the over/under-estimated amount of each galaxy subsample can be determined proportional to the total over/under-estimated amount of sources.

\begin{figure}
\centering
\includegraphics[angle=90,width=1.0\columnwidth]{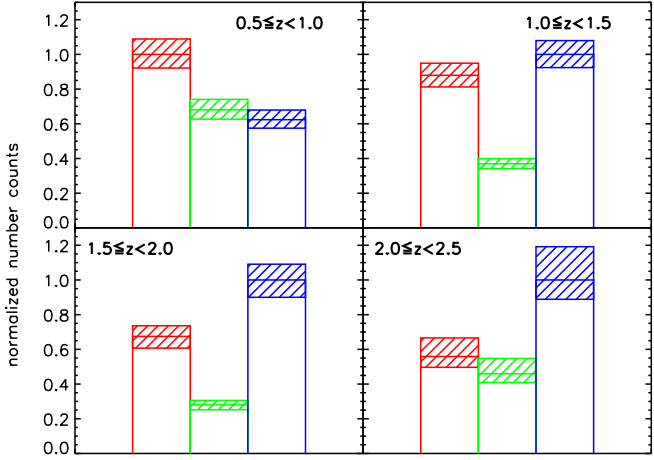}
\caption{Number distribution of COSMOS/UltraVISTA galaxies. The number counts of RS, GV and BC galaxies in different redshift bins are shown as the red, green and blue open histograms respectively, with the maximum number counts in each redshift bin normalized to 1. The line-filled parts in the histograms show the fluctuations in number counts due to selection effect for all galaxy subsamples.}
\label{fig:se_effect}
\end{figure}

Figure~\ref{fig:se_effect} shows the distributions of number counts of all galaxy subsamples selected with $\log(M_\ast/M_\odot)\ge10.0$ as well as their 1$\sigma$ uncertainties caused by selection effect, with the maximum number at each redshift normalized to 1.
At $z<2$, the fluctuations in the number counts of all galaxy subsamples caused by selection effect are generally around 8-9\% of the total subsample source amounts, while it increases to $\sim15\%$ in the highest redshift bin of $2.0\le z\le2.5$.

We account for this source selection uncertainty in the calculation of lifetime error bars for all galaxy subsamples.
The error bar of each subsample is computed as the quadrature sum of the uncertainty caused by selection effect and the uncertainty caused by the measured DM halo mass.
In general, the uncertainty of selection effect is smaller than the uncertainty caused by the measured DM halo mass for each galaxy subsample.
For BC and GV galaxies, the lifetime uncertainties caused by selection effect only contribute to $\approx1/3$ of the total uncertainties.
While for RS galaxies, this fraction is smaller, being about $1/4\sim1/5$.

\begin{figure}
\centering
\includegraphics[width=\columnwidth]{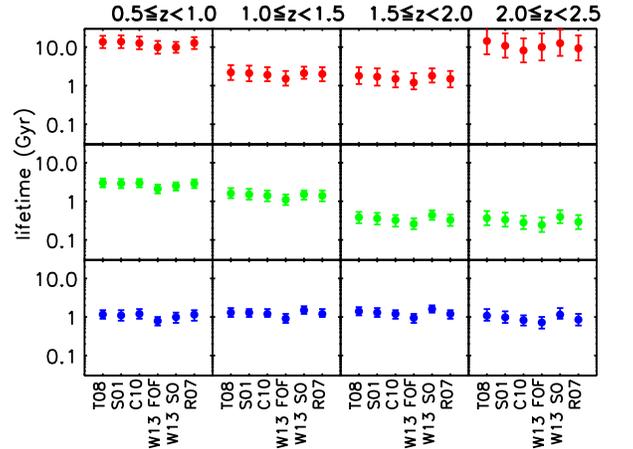}
\caption{Comparison of the lifetimes of UV-selected RS (red symbols in the top row), GV (green symbols in the middle row), and BC galaxies (blue symbols in the bottom row) in four redshift bins calculated using different HMFs. The X-axis shows different HMFs where ``T08'' refers to \citet{Tinker_2008}, ``S01'' refers to \citet{Sheth_2001}, ``C10'' refers to \citet{Crocce_2010}, ``W13 FOF'' and ``W13 SO'' refers to FOF and SO halo-finding methods respectively in \citet{Watson_2013}, and ``R07'' refers to \citet{Reed_2007}.}
\label{fig:lft_HMFs}
\end{figure}

We show the lifetimes of our subsamples calculated with various HMFs in Figure~\ref{fig:lft_HMFs}, and tabulate them in Table~\ref{tab:lft_HMFs}. The variations of lifetime between different HMFs are generally around ${\rm0.2~dex}$ for all galaxy subsamples, with that for the low-redshift subsamples ($z<1.5$) being slightly below ${\rm0.2~dex}$ and that for the high-redshift subsamples ($z\ge1.5$) being slightly above ${\rm0.2~dex}$, respectively. These variations are much smaller than the intrinsic variations of lifetimes of galaxies with different colors at different redshifts.

The RS galaxies generally exhibit longer lifetimes than the BC and GV galaxies at all redshifts, which means that the RS galaxies are more likely in a stable phase. For the BC galaxies, their lifetimes are approximately around ${\rm1~Gyr}$ in all redshift bins, while for the GV galaxies, their lifetimes show a difference between high and low redshift bins.
At $z\ge1.5$, the lifetime of GV galaxies is generally around ${\rm300~Myr}$, and it increases significantly to above ${\rm1~Gyr}$ when the redshift goes to $z<1.5$.
We make a more detailed discussion about the variations of lifetimes among different subsamples in Section~\ref{sect:modeling}.

\begin{table*} \caption{Lifetimes of galaxy subsamples derived from different HMFs in the COSMOS/UltraVISTA field\label{tbl-3}}
\centering

\begin{tabular}{lcccccccccc}
\hline
\hline
Galaxy subsample &
T08 &
S01 &
C10 &
W13 FOF &
W13 SO &
R07 \\
\hline
$0.5 \le z < 1.0$ & \multicolumn{6}{c}{ $\log(M_\ast/M_\odot) \ge 10$ } & \\
$\rm RS$ & $13.7^{+5.9}_{-4.3}$~Gyr & $13.8^{+6.2}_{-4.4}$~Gyr & $12.6^{+6.2}_{-3.8}$~Gyr & $9.9^{+4.5}_{-3.2}$~Gyr & $9.9^{+3.7}_{-2.8}$~Gyr & $12.8^{+5.3}_{-3.9}$~Gyr   \\
$\rm GV$ & $3.0^{+0.9}_{-0.7}$~Gyr & $2.9^{+0.9}_{-0.7}$~Gyr & $3.0^{+0.8}_{-0.7}$~Gyr & $2.1^{+0.6}_{-0.5}$~Gyr & $2.5^{+0.6}_{-0.6}$~Gyr & $2.9^{+0.8}_{-0.7}$~Gyr    \\
$\rm BC$ & $1.2^{+0.3}_{-0.3}$~Gyr & $1.1^{+0.4}_{-0.3}$~Gyr & $1.2^{+0.4}_{-0.3}$~Gyr & $0.8^{+0.2}_{-0.2}$~Gyr & $1.0^{+0.3}_{-0.3}$~Gyr & $1.2^{+0.3}_{-0.4}$~Gyr    \\
\hline
$1.0 \le z < 1.5$ & \multicolumn{6}{c}{ $\log(M_\ast/M_\odot) \ge 10$ } & \\
$\rm RS$ & $2.2^{+1.2}_{-0.8}$~Gyr & $2.1^{+1.2}_{-0.8}$~Gyr & $1.9^{+1.1}_{-0.6}$~Gyr & $1.5^{+0.9}_{-0.5}$~Gyr & $2.1^{+1.0}_{-0.6}$~Gyr & $2.0^{+1.0}_{-0.7}$~Gyr    \\
$\rm GV$ & $1.6^{+0.6}_{-0.4}$~Gyr & $1.5^{+0.6}_{-0.4}$~Gyr & $1.4^{+0.5}_{-0.4}$~Gyr & $1.1^{+0.4}_{-0.3}$~Gyr & $1.5^{+0.4}_{-0.4}$~Gyr & $1.4^{+0.5}_{-0.4}$~Gyr    \\
$\rm BC$ & $1.3^{+0.4}_{-0.3}$~Gyr & $1.3^{+0.3}_{-0.3}$~Gyr & $1.2^{+0.4}_{-0.2}$~Gyr & $0.9^{+0.3}_{-0.2}$~Gyr & $1.5^{+0.4}_{-0.3}$~Gyr & $1.2^{+0.4}_{-0.2}$~Gyr    \\
\hline
$1.5 \le z < 2.0$ & \multicolumn{6}{c}{ $\log(M_\ast/M_\odot) \ge 10$ } & \\
$\rm RS$ & $1.8^{+1.2}_{-0.7}$~Gyr & $1.7^{+1.1}_{-0.7}$~Gyr & $1.5^{+0.8}_{-0.6}$~Gyr & $1.2^{+0.9}_{-0.4}$~Gyr & $1.8^{+1.0}_{-0.6}$~Gyr & $1.5^{+0.9}_{-0.6}$~Gyr    \\
$\rm GV$ & $390^{+140}_{-120}$~Myr & $360^{+140}_{-110}$~Myr & $320^{+120}_{-100}$~Myr & $260^{+100}_{-70}$~Myr & $440^{+140}_{-110}$~Myr & $330^{+120}_{-100}$~Myr    \\
$\rm BC$ & $1.4^{+0.4}_{-0.3}$~Gyr & $1.3^{+0.4}_{-0.3}$~Gyr & $1.2^{+0.3}_{-0.3}$~Gyr & $0.9^{+0.3}_{-0.2}$~Gyr & $1.6^{+0.4}_{-0.3}$~Gyr & $1.2^{+0.3}_{-0.3}$~Gyr    \\
\hline
$2.0 \le z \le 2.5$ & \multicolumn{6}{c}{ $\log(M_\ast/M_\odot) \ge 10$ } &  \\
$\rm RS$ & $14.3^{+17.9}_{-7.8}$~Gyr & $10.8^{+11.9}_{-5.5}$~Gyr & $8.2^{+8.6}_{-4.2}$~Gyr & $10.0^{+12.6}_{-5.5}$~Gyr & $12.6^{+15.7}_{-6.7}$~Gyr & $9.3^{+10.7}_{-4.8}$~Gyr   \\
$\rm GV$ & $370^{+190}_{-130}$~Myr & $330^{+180}_{-110}$~Myr & $280^{+140}_{-90}$~Myr & $240^{+140}_{-80}$~Myr & $400^{+190}_{-130}$~Myr & $290^{+150}_{-100}$~Myr   \\
$\rm BC$ & $1.1^{+0.5}_{-0.3}$~Gyr & $1.0^{+0.4}_{-0.3}$~Gyr & $0.8^{+0.3}_{-0.2}$~Gyr & $0.7^{+0.3}_{-0.2}$~Gyr & $1.2^{+0.5}_{-0.3}$~Gyr & $0.9^{+0.3}_{-0.3}$~Gyr   \\
\hline
\end{tabular}

\label{tab:lft_HMFs}
\end{table*}

\subsection{Influence of Redshift Bin Width}\label{sect:dis_zbins}

The width of redshift bin can also affect the resultant lifetime of a galaxy sample.
This is because if we choose a very narrow redshift bin, our assumption on the DM halo experiencing a specific phase of galaxy evolution would no longer hold.
Thus it is necessary to know, to which extent the width of redshift bin can affect the lifetime estimation.

\begin{figure*}
\centering
\includegraphics[ width=0.49\textwidth]{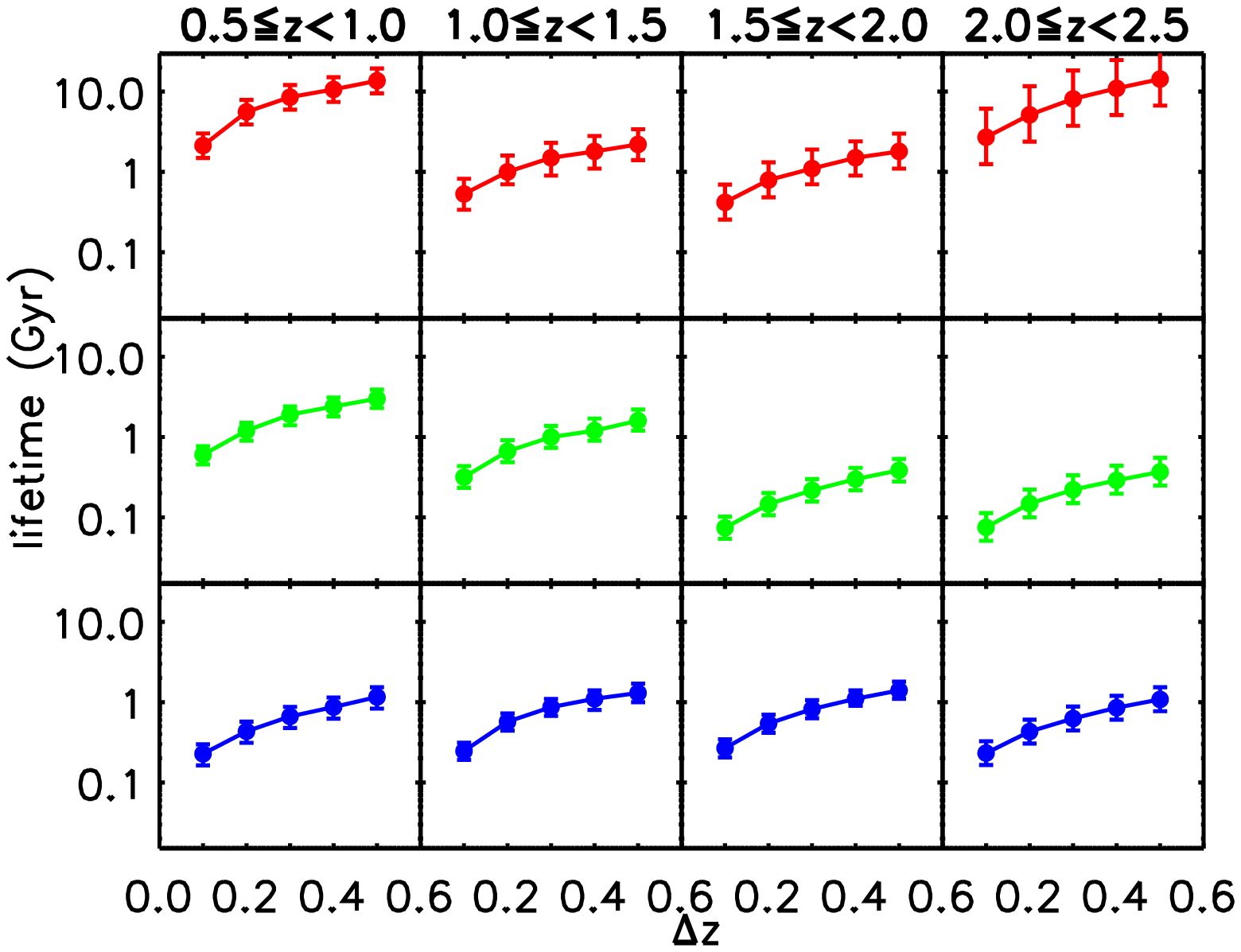}
\includegraphics[ width=0.49\textwidth]{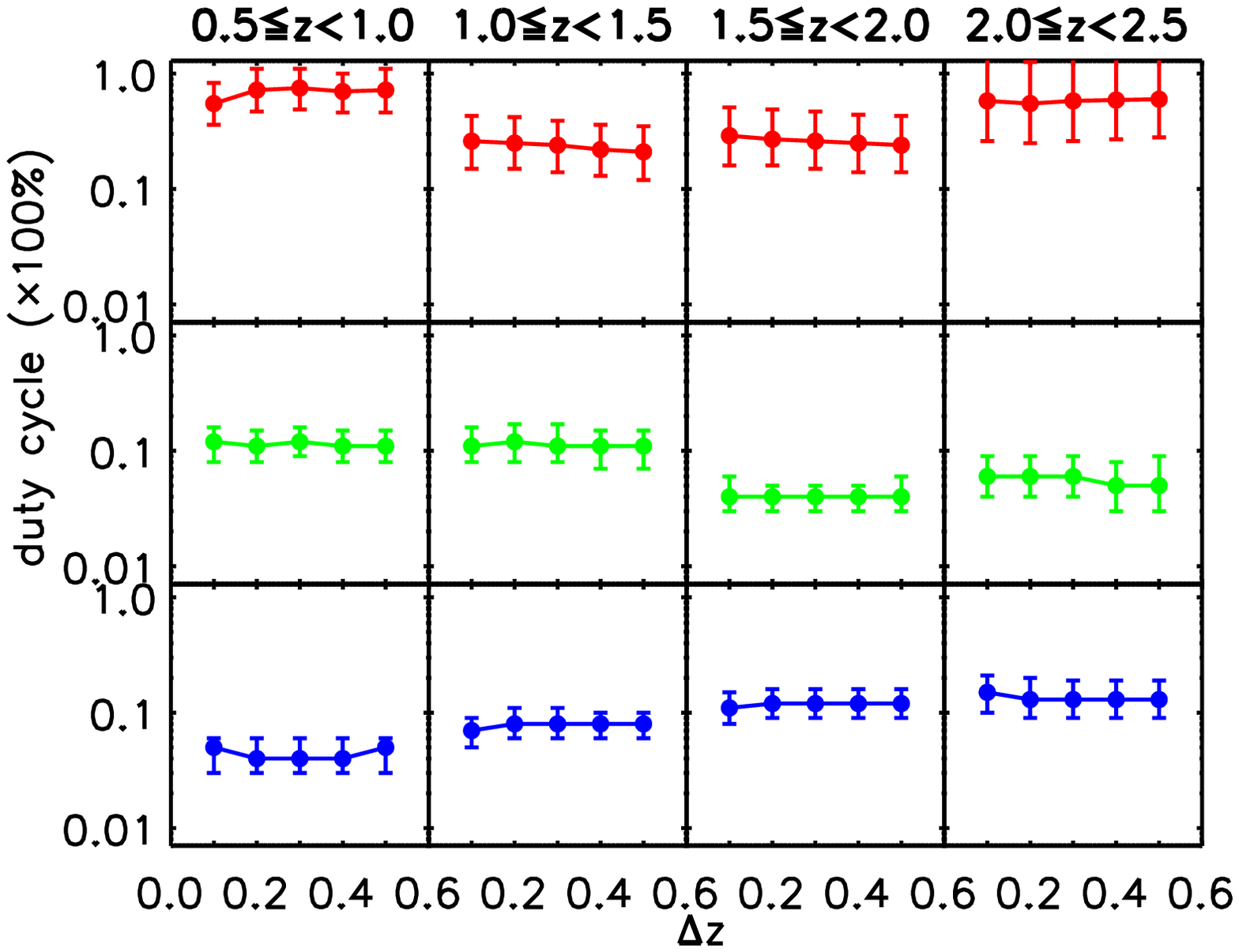}
\caption{The lifetimes (left panels) and the corresponding duty cycles (right panels) of the RS (red symbols in the top row), GV (green symbols in the middle row), and BC (blue symbols in the bottom row) galaxies as a function of the redshift bin width, centering at the median of each redshift bin annotated at the top of each column.
}
\label{fig:lft_zbins}
\end{figure*}

We resample each of our galaxy populations with finer redshift bin widths of $\Delta z=0.1,0.2,0.3,0.4,0.5$, centering at $\bar{z}=0.75,1.25,1.75,2.25$.
Here, we only apply the \citet{Tinker_2008} HMF and do not consider the change of DM halo mass within the redshift bin since we have checked, in \citet{Lin_2019}, that the halo mass does not show an obvious evolution within each redshift bin. The results are shown in Figure~\ref{fig:lft_zbins} and tabulated in Table~\ref{tab:lft_zbins}.

\begin{table*} \caption{Lifetimes of galaxy subsamples derived by different redshift bin widths in the COSMOS/UltraVISTA field\label{tbl-4}}
\centering

\begin{tabular}{lcccccccccc}
\hline
\hline
Galaxy subsample &
$\Delta z=0.1$ &
$\Delta z=0.2$ &
$\Delta z=0.3$ &
$\Delta z=0.4$ &
$\Delta z=0.5$ & \\
\hline
$0.5 \le z < 1.0$ & \multicolumn{5}{c}{ $\log(M_\ast/M_\odot) \ge 10$ } & \\
$\rm RS$ & $2.1^{+0.9}_{-0.6}$~Gyr & $5.6^{+2.3}_{-1.7}$~Gyr & $8.5^{+3.6}_{-2.5}$~Gyr & $10.6^{+4.5}_{-3.2}$~Gyr & $13.7^{+5.7}_{-4.2}$~Gyr   \\
$\rm GV$ & $0.6^{+0.2}_{-0.1}$~Gyr & $1.2^{+0.3}_{-0.3}$~Gyr & $1.9^{+0.5}_{-0.5}$~Gyr & $2.4^{+0.7}_{-0.6}$~Gyr & $3.0^{+0.9}_{-0.7}$~Gyr    \\
$\rm BC$ & $0.2^{+0.1}_{-0.1}$~Gyr & $0.4^{+0.2}_{-0.1}$~Gyr & $0.7^{+0.2}_{-0.2}$~Gyr & $0.9^{+0.2}_{-0.3}$~Gyr & $1.2^{+0.3}_{-0.4}$~Gyr    \\
\hline
$1.0 \le z < 1.5$ & \multicolumn{5}{c}{ $\log(M_\ast/M_\odot) \ge 10$ } & \\
$\rm RS$ & $0.5^{+0.3}_{-0.2}$~Gyr & $1.0^{+0.6}_{-0.3}$~Gyr & $1.5^{+0.8}_{-0.6}$~Gyr & $1.8^{+1.0}_{-0.7}$~Gyr & $2.2^{+1.2}_{-0.8}$~Gyr    \\
$\rm GV$ & $0.3^{+0.1}_{-0.1}$~Gyr & $0.7^{+0.2}_{-0.2}$~Gyr & $1.0^{+0.4}_{-0.3}$~Gyr & $1.2^{+0.5}_{-0.3}$~Gyr & $1.6^{+0.6}_{-0.4}$~Gyr    \\
$\rm BC$ & $0.2^{+0.1}_{-0.1}$~Gyr & $0.6^{+0.1}_{-0.1}$~Gyr & $0.9^{+0.2}_{-0.2}$~Gyr & $1.1^{+0.3}_{-0.3}$~Gyr & $1.3^{+0.4}_{-0.3}$~Gyr    \\
\hline
$1.5 \le z < 2.0$ & \multicolumn{5}{c}{ $\log(M_\ast/M_\odot) \ge 10$ } & \\
$\rm RS$ & $0.4^{+0.3}_{-0.1}$~Gyr & $0.8^{+0.4}_{-0.3}$~Gyr & $1.1^{+0.8}_{-0.4}$~Gyr & $1.5^{+0.9}_{-0.6}$~Gyr & $1.8^{+1.2}_{-0.7}$~Gyr    \\
$\rm GV$ & $70^{+30}_{-20}$~Myr & $150^{+50}_{-40}$~Myr & $220^{+80}_{-60}$~Myr & $300^{+110}_{-80}$~Myr & $390^{+140}_{-110}$~Myr    \\
$\rm BC$ & $0.3^{+0.1}_{-0.1}$~Gyr & $0.5^{+0.2}_{-0.1}$~Gyr & $0.8^{+0.3}_{-0.2}$~Gyr & $1.1^{+0.3}_{-0.2}$~Gyr & $1.4^{+0.4}_{-0.3}$~Gyr    \\
\hline
$2.0 \le z \le 2.5$ & \multicolumn{5}{c}{ $\log(M_\ast/M_\odot) \ge 10$ } &  \\
$\rm RS$ & $2.7^{+3.4}_{-1.4}$~Gyr & $5.1^{+6.6}_{-2.7}$~Gyr & $8.1^{+10.2}_{-4.3}$~Gyr & $11.0^{+13.7}_{-5.9}$~Gyr & $14.3^{+17.7}_{-7.6}$~Gyr   \\
$\rm GV$ & $80^{+30}_{-30}$~Myr & $150^{+70}_{-50}$~Myr & $220^{+110}_{-70}$~Myr & $290^{+150}_{-90}$~Myr & $370^{+180}_{-120}$~Myr   \\
$\rm BC$ & $0.2^{+0.1}_{-0.1}$~Gyr & $0.4^{+0.2}_{-0.1}$~Gyr & $0.6^{+0.3}_{-0.2}$~Gyr & $0.9^{+0.3}_{-0.3}$~Gyr & $1.1^{+0.4}_{-0.3}$~Gyr   \\
\hline
\end{tabular}

\label{tab:lft_zbins}
\end{table*}

The lifetimes of all galaxy subsamples at each redshift generally increase with the increasing redshift bin width, $\Delta z$. But the increasing trend becomes flatter when the bin width is large enough, and the lifetime of each subsample almost approaches a constant value when $\Delta z$ reaches 0.5. Therefore, we conclude that $\Delta z=0.5$ is large enough to correctly estimate the lifetime of each UV-selected galaxy subsample.

We note that although the lifetime of each subsample shows some dependence on the choice of redshift bin width, its duty cycle remains roughly unchanged with $\Delta z$, as is shown in the right panels of Figure~\ref{fig:lft_zbins}.
We define the duty cycle of a subsample as the fraction of DM halos hosting the respective type of galaxies within its redshift interval, following the definition in e.g., \citet{Martini_2001},\citet{Haiman_2001},\citet{Hickox_2012},\citet{Eftekharzadeh_2015},\citet{Toba_2017}, and \citet{He_2017}:
\begin{equation}
f_{\rm duty} = \frac{n_{\rm sample}}{\int^{\infty}_{M_{\rm h,min}}\frac{dn}{dM_{\rm h}}dM_{\rm h}},
\end{equation}
where $n_{\rm sample}$ is the number density of each subsample, $\frac{dn}{dM_{\rm h}}$ is the HMF following the functional form of \citet{Tinker_2008}, and $M_{\rm h,min}$ is adopted as the estimated lower limit of the DM halo mass of the subsample, respectively.

It is clear that BC and GV galaxies generally have smaller duty cycles than RS galaxies at the same redshifts. And for BC and GV galaxies, their duty cycles show distinct behaviors as the redshift evolves. The duty cycle of BC galaxies increases with the increasing redshift, from 4\% at $z\sim0.7$ to 13\% at $z\sim2.2$; while that of GV galaxies shows an opposite trend, decreasing from 11\% at $z\sim0.7$ to 6\% at $z\sim2.2$.

In addition, it is interesting to find that the duty cycles of BC and GV galaxies are comparable to that of AGN at corresponding redshifts.
\citet{Shankar_2009} constructed evolutionary models to estimate the properties of AGNs and found that their duty cycle shows a declination from 7\% at $z=3$ to 0.4\% at $z=1$, when the black hole mass is fixed to $10^9~M_\odot$.
Some other studies also utilized the same clustering estimator to evaluate the duty cycles of AGNs \citep[e.g.,][]{White_2008,Gilli_2009,Shankar_2010}, implying a result of $\sim10\%$ on average.
And it has also been discovered that the duty cycle of AGNs is sensitively dependent on the luminosity, as well as other cosmological parameters (e.g., $\sigma_8$) and the choice of HMF \citep[e.g.,][]{Adelberger_2005,Shen_2007,Eftekharzadeh_2015,He_2017}. More luminous AGNs usually tend to have shorter lifetimes and smaller duty cycles.

\subsection{Influence of DM Halo Growth}\label{sect:dis_halogr}

\begin{figure*}
\centering
\includegraphics[ width=0.49\textwidth]{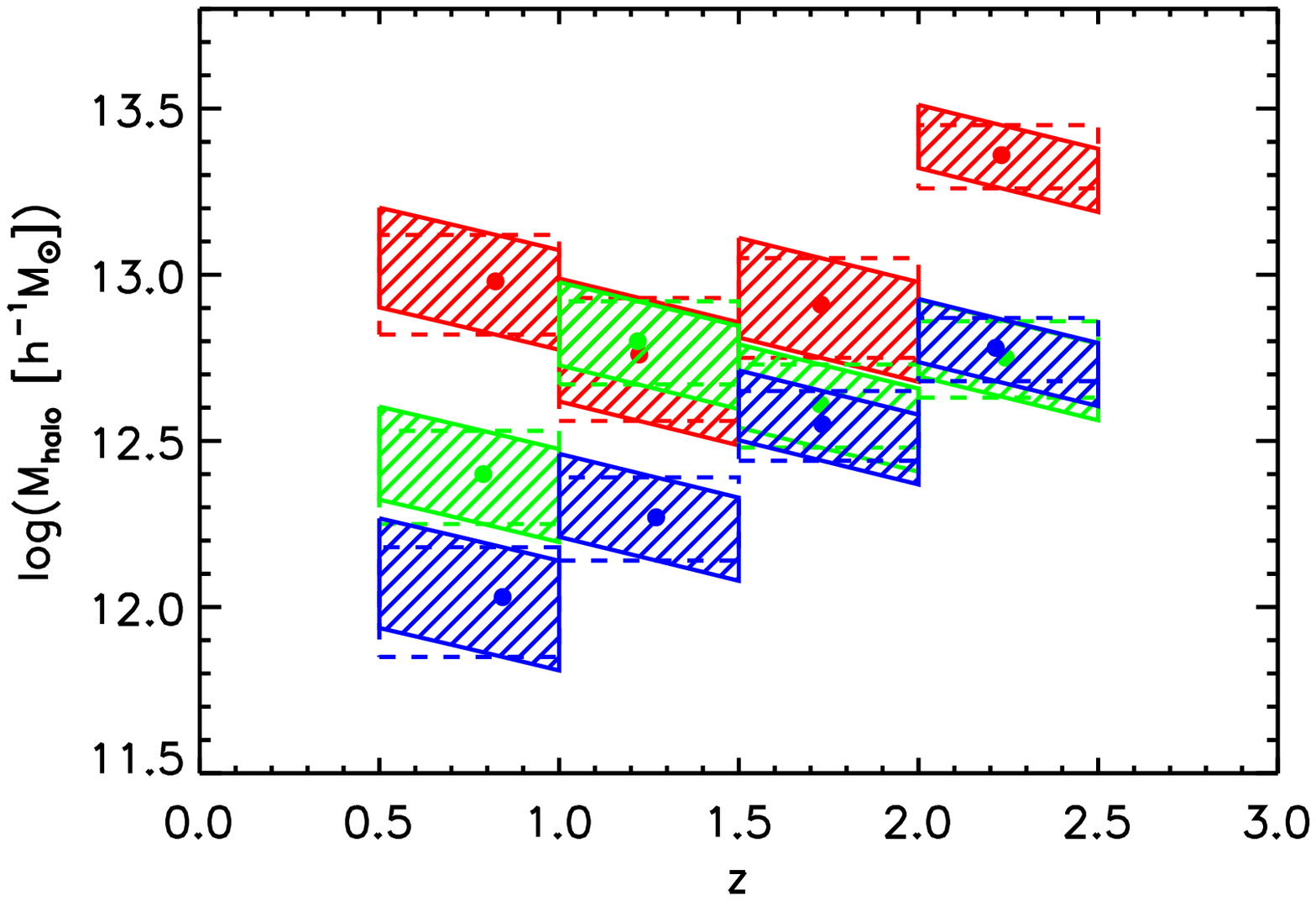}
\includegraphics[ width=0.49\textwidth]{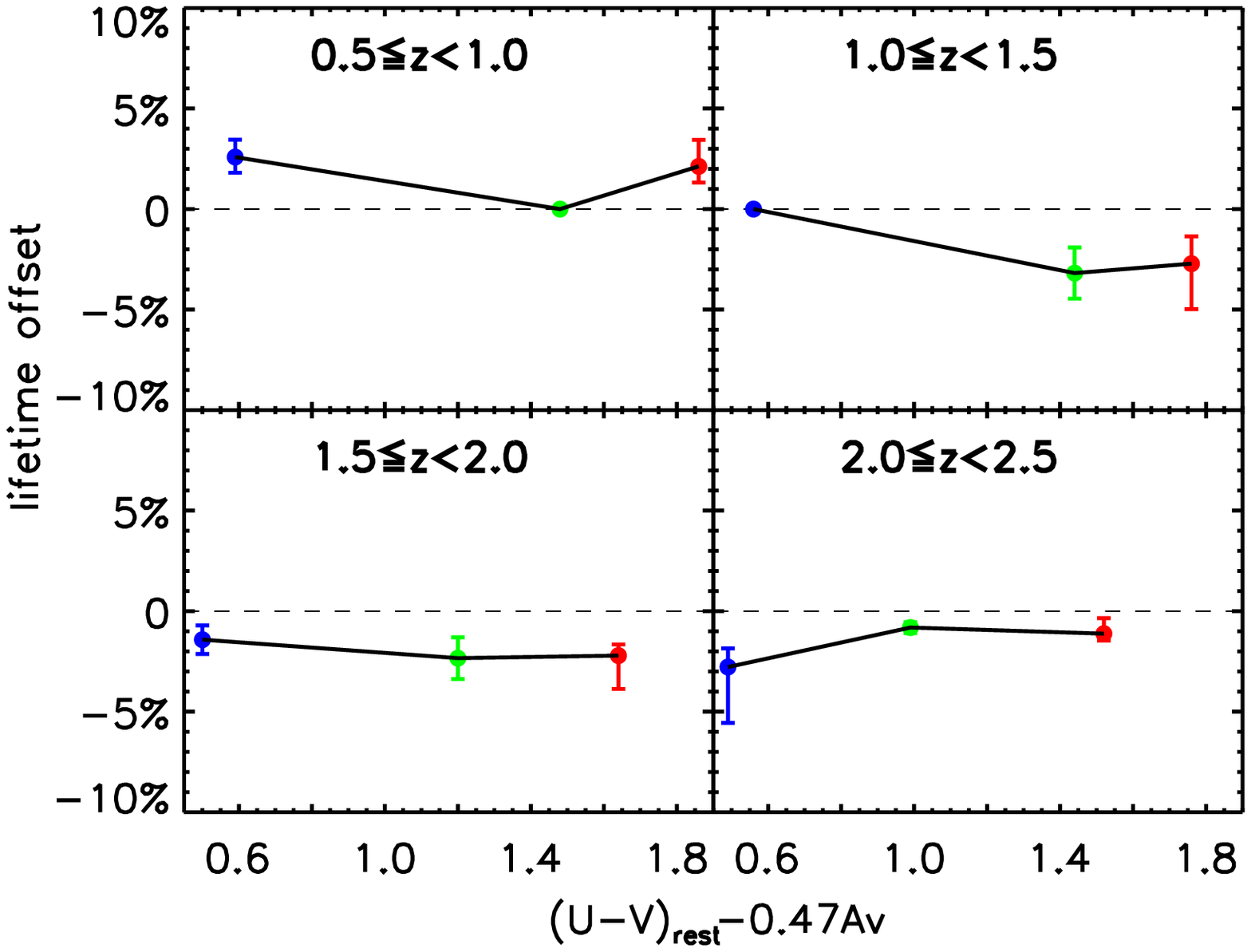}
\caption{{The left panel} shows the integral intervals of the HMF in Equation~\ref{eq:integral_Nhalo} as a function of redshift for each of the UV-selected galaxy subsample with $\log(M_{\ast}/M_{\odot})\ge10.0$. The original integral intervals are shown as the dashed open rectangles, while the integral intervals considering the growth of DM halo mass are shown as the solid line-filled polygons. The filled circles show the halo masses of these galaxy subsamples at their respective median redshifts. {The right panel} shows the lifetime offset as a function of the extinction-corrected UV color.
}
\label{fig:lft_halogr}
\end{figure*}

As stated in the Introduction, the activity of galaxies is usually accompanied with the growth of their host DM halos, therefore it is of necessity to know whether and to which extent the growth of DM halos would affect the lifetimes of our galaxy subsamples.
We assume the growth of DM halos following the median mass growth rate of DM halos derived by \citet{Fakhouri_2010} through fitting the evolution of DM halos in the Millennium-II cosmology simulation. The growth of DM halos will change the integral interval in Equation~\ref{eq:integral_Nhalo}, since now we need to redefine the DM halos with a mass similar to that of the galaxy subsamples.
With the growing DM halo mass, the upper and lower limits of the halo mass of each galaxy subsample would also increase in time following the growth function by \citet{Fakhouri_2010}, thus the integral interval of halo mass is redshift-dependent within each redshift bin.
The original integral intervals as well as those considering the growth of DM halos are shown as the dashed open and solid line-filled polygons in the left panel of Figure~\ref{fig:lft_halogr}, respectively.

For each of our galaxy subsample selected by $\log(M_{\ast}/M_{\odot})\ge10.0$, we evaluate the offset of the lifetime accounting for the growth of DM halos. The offset is defined as
\begin{equation}
lifetime~offset = \frac{lifetime_{\rm halo~growth}-lifetime}{lifetime},
\end{equation}
where $lifetime_{\rm halo~growth}$ is the lifetime considering the growth of DM halos. We apply the \citet{Tinker_2008} HMF during this computation.

The result is shown in the right panel of Figure~\ref{fig:lft_halogr}.
For all of our subsamples, the lifetime offsets are no more than 5\%, considering the median growth rate of DM halos by \citet{Fakhouri_2010}. Thus we conclude that, the growth of DM halo would have a negligible effect on the lifetimes of our UV-selected galaxies in these redshift bins.

\subsection{Influence of Stellar Mass}\label{sect:dis_smass}

\begin{figure}
\centering
\includegraphics[width=\columnwidth]{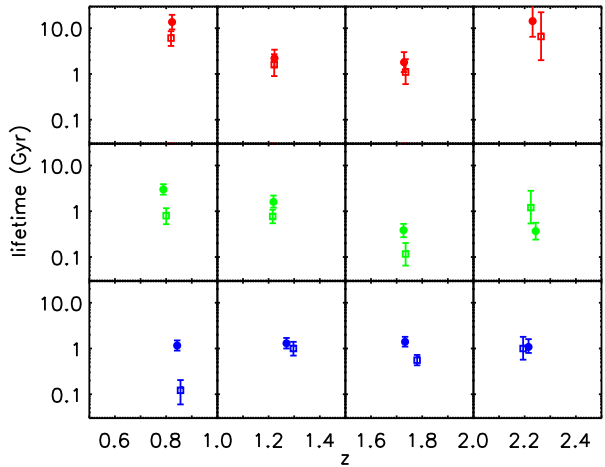}
\caption{Comparison of the lifetimes of galaxy subsamples selected by different stellar-mass limits in the COSMOS/UltraVISTA field.
The filled circles, open rectangles, and open triangles represent the lifetimes of galaxy subsamples selected by ${\log(M_\ast/M_\odot)\ge10.0}$, 10.5, and 11.0, respectively.
For galaxy subsamples with the same colors and in the same redshift range, the derived lifetime increases with the decreasing stellar-mass limit; however, this difference diminishes with the increasing redshift.
 }
\label{fig:lft_smass}
\end{figure}

We also compare the lifetimes of galaxy subsamples selected by different stellar-mass limits {\bf $\log(M_\ast/M_\odot)\ge10.5$} as tabulated in Table~\ref{tab:galaxy_bin}. For all subsamples selected {with a higher} stellar-mass threshold, we calculate their lifetimes using the \citet{Tinker_2008} HMF and show the results in Figure~\ref{fig:lft_smass}.
We have also accounted for source bias and selection effect similarly to what we have done in the lifetime calculation for the subsamples selected with $\log(M_\ast/M_\odot)\ge10.0$.

Generally speaking, for a given color of galaxies, the subsample selected with a larger stellar-mass cut would have a shorter lifetime at a given redshift.
In the low-redshift bins this trend is obvious, but it seems to diminish when the redshift goes higher.

The lifetimes of different galaxy populations, which characterize the time when a certain type of galaxies stay in this phase, can also reflect the typical quenching timescale of SFGs when we consider the quenching scenario. In this scenario, due to mechanisms such as AGN feedback \citep[e.g.,][]{Bower_2006,Cicone_2014,Schawinski_2014} or merger-driven disk instability \citep[e.g.,][]{Martig_2009,Leauthaud_2012,Zolotov_2015}, the high-redshift SFGs would consume their gas and transform into QGs. Multiple studies have revealed that this quenching process is much more efficient in more massive galaxies \citep[e.g.,][]{Springel_2005,Zolotov_2015,Contini_2020}.
But at higher redshifts, the difference between the quenching timescales of galaxies with different stellar-mass limits seems to be diluted, as it is evident from some studies on the galaxy mass function (e.g., \citealt{Ilbert_2013}), that over the entire stellar mass range, the number density of high-redshift ($z>1$) QGs increases smoothly.
It is also supported by some other studies on the emergence of QGs that the quenching timescale of high-redshift compact SFGs is very short, which can be as short as several hundred million years \citep[e.g.,][]{Barro_2013,Barro_2016,Kocevski_2017}. And this quenching timescale seems to be even shorter with the increasing redshift, which was proposed by some studies on the luminosity function \citep[e.g.,][]{Cucciati_2012} and the stellar-mass function \citep[e.g.,][]{Rowlands_2018} of SFGs. Thus it is difficult to distinguish the differences of the SFR transition timescales between galaxies with different stellar masses at high redshifts.

\section{Comparison with the SED Modeling}\label{sect:modeling}

In this section, we compare the galaxy lifetimes with those derived from the SED modeling.
In section~\ref{sect:sed_modeling}, we introduce the method of our SED modeling and discuss about the evolutionary track degeneracy due to the choice of different sets of parameters.
In section~\ref{sect:lft_compare}, we make a comparison between the SED modeled lifetimes and the aforementioned clustering-based lifetimes. We also compare the lifetimes between the populations of different colors and their evolutions as a function of redshift.
In section~\ref{sect:lft_GV_short}, we discuss the possible reasons that cause the short lifetime of GV galaxies at $z\ge1.5$.
In section~\ref{sect:gas}, we compare the lifetimes of the subsamples with their typical gas depletion timescales.
In section~\ref{sect:reproduce}, we try to reproduce the photometric redshift distribution of all UV-selected galaxies in the COSMOS/UltraVISTA catalog \citep{Muzzin_2013a} based on the derived lifetimes of this paper and a preset HMF, and we also compare the best parameters in reproducing the galaxy redshift distributions with our lifetime results.

\subsection{The SED modeling}\label{sect:sed_modeling}

\begin{figure}
\centering
\includegraphics[width=0.8\columnwidth, angle=90]{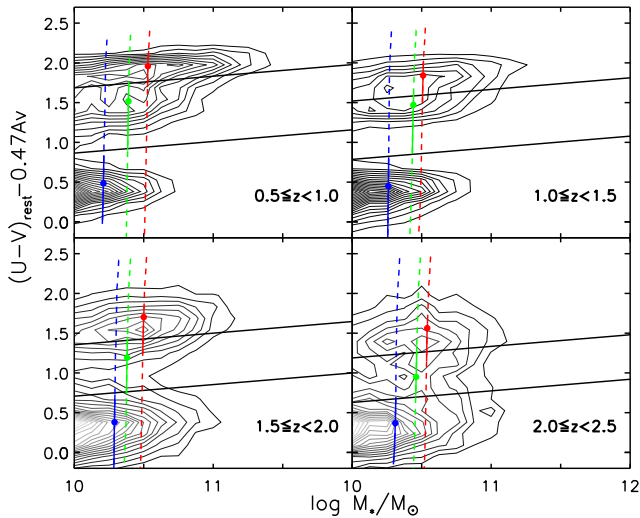}
\caption{The color-stellar mass diagram for the UV-selected galaxies with ${\log(M_\ast/M_\odot)\ge10.0}$ and $0.5\le z\le2.5$ (contours). Black lines show the rough galaxy separation criteria fixing the redshift in Equation~\ref{ref:equ_sep_cri} to the median of each redshift bin. The modeled evolution tracks on the color-mass diagram for different galaxy populations are shown as the lines of different colors with the filled circles representing the median stellar masses and dust-corrected UV colors of different subsamples. The solid portion of each line shows the modeled lifetime of each sample, which is defined as the period of the track on the color-mass diagram that satisfies the separation criteria of Equation~\ref{ref:equ_sep_cri} for each galaxy population.}
\label{fig:cmr_evo}
\end{figure}

We utilize a simple stellar population generated from the \citet{BC03} stellar population synthesis library assuming a \citet{Chabrier_2003} initial mass function and solar metallicity. To construct the SED templates, we further assume exponentially-declining star-formation histories with the form of ${{\rm SFR}\propto \exp(-t/\tau)}$, where $t$ is the time since the onset of star formation, and $\tau$ is the e-folding timescale, respectively. We do not take dust extinction into account in this SED modeling, since we have divided our galaxy sample using the dust-corrected rest-frame UV colors.
We generate galaxy SEDs formed at different redshifts between $z=0-10$ and the e-folding timescale $\tau$ is allowed to vary between $10^7~\rm{yr}$ and $10^{10}~\rm{yr}$.
The dust-corrected UV colors of a modeled SED at different redshifts are computed through the convolution between the redshifted galaxy SED and the UV filters.
The evolution track of each galaxy subsample is determined by finding the right one whose dust-corrected UV color ($\rm{UV_{rest,ext-cor}}$), stellar mass and e-folding timescale $\tau$ match those respective median SED-fitting values of each galaxy sample at its median redshift in the \citet{Muzzin_2013a} catalog.
The modeled evolution tracks of color-stellar mass relation for all UV-selected galaxy subsamples with ${\log(M_\ast/M_\odot)\ge10.0}$ are shown as solid color lines in Figure~\ref{fig:cmr_evo}.

We define the modeled lifetime as the period when the dust-corrected UV color and stellar mass of the modeled evolution track of each galaxy subsample in the color-mass diagram satisfy the separation criteria of the respective galaxy population in Equation~\ref{ref:equ_sep_cri}, shown as the solid portions of the lines in Figure~\ref{fig:cmr_evo}. Specifically, we define the modeled lifetime of the BC, GV, or RS galaxies as the period when the respective color-mass relation computed from the modeled SED agrees with that of BC, GV, or RS galaxies since the formation redshift $z_f$. In addition, for the modeled lifetime of each RS galaxy population, we restrict the redshift to be no smaller than the observed lower redshift limit of each subsample during the evolution. But if we consider the red sequence as a stable phase in the quenching scenario and neglect other factors such as rejuvenation for a small amount of red dead galaxies  \citep[e.g.,][]{Reipurth_1997,Portegies_2000,Perets_2009}, the expected lifetime of the RS galaxies can be much longer than what we model. The modeled lifetimes of all galaxy populations together with those inferred through clustering analyses are all tabulated in Table~\ref{tab:lft_model}.

\begin{figure*}
\centering
\includegraphics[ width=0.49\textwidth]{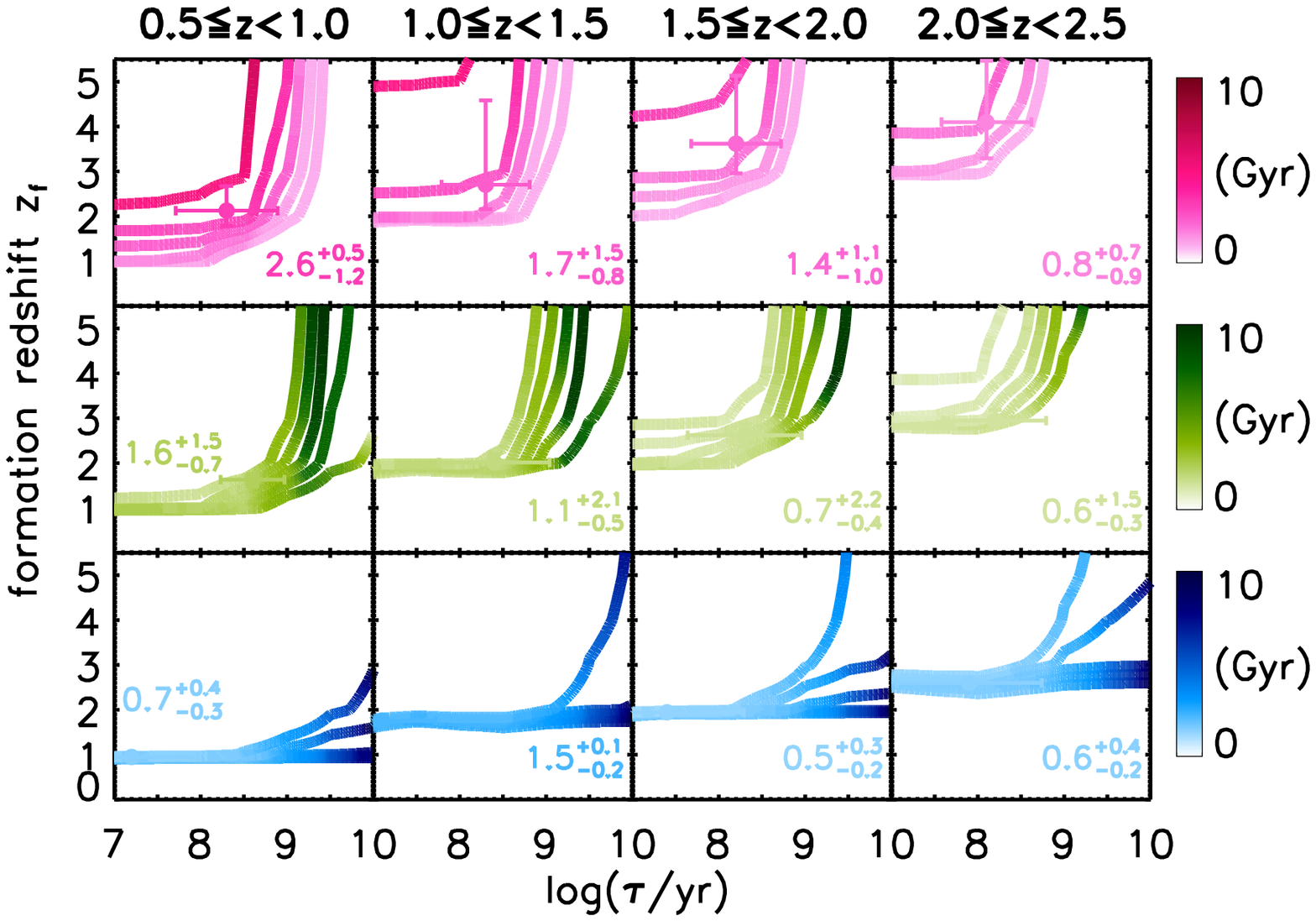}
\includegraphics[ width=0.49\textwidth]{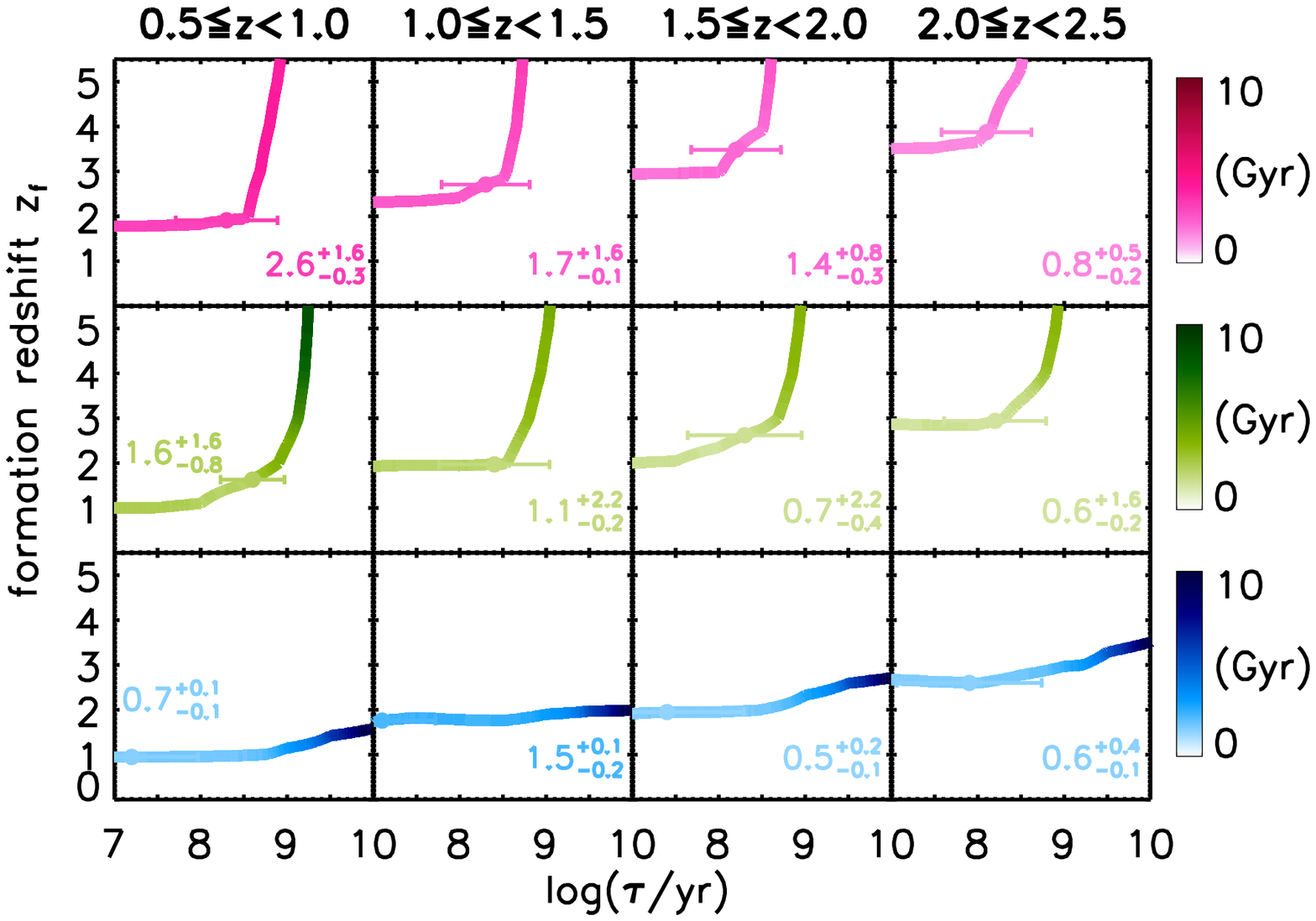}
\caption{ The degeneracy of evolutionary track in the color-mass diagram due to multiple choices of parameter pairs of $z_f$ and $\tau$. The left panel shows the degeneracy of $z_f$ and $\tau$ by fixing the normalized stellar mass to the median $M_\ast$ and allowing $(U-V)_{\rm rest}-0.47A_v$ to vary in the step of 0.2 (RS: $[1.2, 2.0]$, GV: $[0.8, 1.6]$, BC: $[0.0, 0.8]$). Each point of the parameter pair of $z_f$ and $\tau$ in the figure corresponds to an evolutionary track running through the median $M_\ast$ and $(U-V)_{\rm rest}-0.47A_v$ at the respective redshift of each subsample. The points are color coded by their modeled lifetimes according to the color bars on the right. Error bars show the $1\sigma$ distributions of $\tau$ and $(U-V)_{\rm rest}-0.47A_v$ around their medians for each subsample. The modeled lifetime of each subsample and the $1\sigma$ uncertainties (see the text for detail) are labeled as color numbers in each sub-panel. The right panel shows the degeneracy of $z_f$ and $\tau$ by fixing the extinction-corrected rest-frame UV color to the median $(U-V)_{\rm rest}-0.47A_v$ of each subsample and allowing the stellar mass to vary between $\log(M_\ast/M_\odot)=10.0$ and $\log(M_\ast/M_\odot)=10.8$. We have also labeled the parameter pair of $z_f$ and $\tau$ corresponding to the evolutionary track running through the median $M_\ast$ and $(U-V)_{\rm rest}-0.47A_v$ of each subsample in each sub-panel, as well as their $1\sigma$ distributions. Color numbers in each sub-panel show the modeled lifetimes and their $1\sigma$ uncertainties.
}
\label{fig:track_dege}
\end{figure*}

Given that we use a single initial mass function with a fixed metallicity and assume a uniform star-formation history, the evolutionary tracks in the color-mass diagram (Figure~\ref{fig:cmr_evo}) are subject to track degeneracy.
The most prominent degeneracy comes from the choices of formation redshift $z_f$ and the exponential declination timescale $\tau$ of the modeled star-formation history, which can reproduce almost the same track across the median $M_\ast$ and $(U-V)_{\rm rest}-0.47A_v$ for each observed galaxy subsample.
Thus we study the choices of $z_f$ and $\tau$ that can produce tracks across each pair of $M_\ast$ and $(U-V)_{\rm rest}-0.47A_v$ in the color-mass diagram.
We do not consider the variations of initial mass function and metallicity in this paper.
We show our results in Figure~\ref{fig:track_dege}.

In the left part of Figure~\ref{fig:track_dege}, we study track degeneracy by fixing the stellar mass to the median $M_\ast$ of each observed subsample selected with $\log(M_\ast/M_\odot)\ge10.0$ and allowing rest-frame extinction-corrected UV color to vary in the step of 0.2.
For the modeling of RS galaxies, the $(U-V)_{\rm rest}-0.47A_v$ values vary from 1.2 to 2.0 shown as five different red curves in each sub-panel of the left part of Figure~\ref{fig:track_dege}.
For the modeling of GV and BC galaxies, the $(U-V)_{\rm rest}-0.47A_v$ values are set to vary within $[0.8, 1.6]$ and $[0.0, 0.8]$ respectively, also shown as color-coded curves in the sub-panels.
Each point on the color-coded curves represents a pair of choices of formation redshift $z_f$ and the exponential star-formation declination timescale $\tau$, corresponding to one of the modeled evolutionary tracks in the color-mass diagram.
The points along the curves are coded with the color gradients according to the color bars of the SED modeled lifetimes on the right hand side.
In each sub-panel, we have also added the modeled parameter pair of $z_f$ and $\tau$ corresponding to the evolutionary track running across the median $M_\ast$ and $(U-V)_{\rm rest}-0.47A_v$ of each subsample, shown as a filled dot with its color coded as its corresponding lifetime.
The horizontal and vertical error bars in each sub-panel denote the $1\sigma$ uncertainties of the exponential star-formation declination timescale $\tau$ and extinction-corrected rest-frame UV color around the median values of each subsample.
The numbers annotated in the sub-panels show the exact modeled lifetimes of subsamples as well as their uncertainties with respect to the maximum and minimum values of the modeled lifetimes in the $1\sigma$ parameter box of $\tau$ and $(U-V)_{\rm rest}-0.47A_v$.

Although there are many parameter pairs that may produce almost the same evolutionary track crossing the median $M_\ast$ and $(U-V)_{\rm rest}-0.47A_v$ of each subsample in the color-mass diagram, the estimated lifetime does not seem to exhibit a large variation within the $1\sigma$ range of $\tau$ and $(U-V)_{\rm rest}-0.47A_v$.
The SED modeled lifetimes of RS galaxies are generally larger than that of BC and GV galaxies at all redshifts.
For GV galaxies, they have similar modeled lifetimes to that of BC galaxies at higher redshifts ($z>1$).
But at lower redshifts ($z\le1$), the modeled lifetimes of GV galaxies increase rapidly, while that of BC galaxies remain at a low level.

We have also studied track degeneracy by fixing the rest-frame extinction-corrected UV color to the median $(U-V)_{\rm rest}-0.47A_v$ of each subsample, and allowing the median normalized stellar mass to vary in the step of 0.2 between $\log(M_\ast/M_\odot)=10.0$ and $\log(M_\ast/M_\odot)=10.8$.
But this time we do not find a clear division of parameter pairs of $z_f$ and $\tau$ between different normalized stellar masses, as can be seen in the right part of Figure~\ref{fig:track_dege}.
This is because when we model the color evolution of a galaxy SED, the stellar mass is a flexible parameter which is independent of the color evolution.
The normalization of stellar mass only depends on the exact value of the median $M_\ast$ of each subsample at the respective redshift.
We also label the modeled lifetimes of each subsample together with their $1\sigma$ uncertainties corresponding to the maximum and minimum modeled lifetimes within the $1\sigma$ distributions of $\tau$ and normalized $M_\ast$.

\begin{table*} \caption{Comparison of the clustering-based Lifetimes with those from the SED modeling \label{tbl-5}}
\centering

\begin{tabular}{lcccccccccc}
\hline
\hline
Galaxy subsample &
$\rm{lifetime~(clustering)^a}$ &
$\rm{lifetime~(SED~modeling)^b}$ &
$\rm{formation~redshift~(z_f)^c}$ &
$\rm{lifetime~(simulation~peak)^d}$ & \\
\hline
$0.5 \le z < 1.0$ & \multicolumn{5}{c}{ $\log(M_\ast/M_\odot) \ge 10$ } & \\
$\rm RS$ & $12.1^{+5.0}_{-3.6}$~Gyr & 2.6~Gyr & 1.91 & 2.8~Gyr   \\
$\rm GV$ & $2.7^{+0.8}_{-0.6}$~Gyr & 1.6~Gyr & 1.63 & 1.9~Gyr    \\
$\rm BC$ & $1.1^{+0.3}_{-0.3}$~Gyr & 0.7~Gyr & 0.95 & 1.1~Gyr    \\
\hline
$1.0 \le z < 1.5$ & \multicolumn{5}{c}{ $\log(M_\ast/M_\odot) \ge 10$ } & \\
$\rm RS$ & $2.0^{+1.0}_{-0.7}$~Gyr & 1.7~Gyr & 2.71 & 1.8~Gyr    \\
$\rm GV$ & $1.4^{+0.5}_{-0.4}$~Gyr & 1.1~Gyr & 1.97 & 1.2~Gyr    \\
$\rm BC$ & $1.3^{+0.3}_{-0.3}$~Gyr & 1.5~Gyr & 1.76 & 1.4~Gyr    \\
\hline
$1.5 \le z < 2.0$ & \multicolumn{5}{c}{ $\log(M_\ast/M_\odot) \ge 10$ } & \\
$\rm RS$ & $1.6^{+1.0}_{-0.6}$~Gyr & 1.4~Gyr & 3.48 & 1.3~Gyr    \\
$\rm GV$ & $350^{+120}_{-90}$~Myr & 780~Myr & 2.62 & 970~Myr    \\
$\rm BC$ & $1.3^{+0.3}_{-0.3}$~Gyr & 0.6~Gyr & 1.95 & 0.8~Gyr    \\
\hline
$2.0 \le z \le 2.5$ & \multicolumn{5}{c}{ $\log(M_\ast/M_\odot) \ge 10$ } &  \\
$\rm RS$ & $10.9^{+12.7}_{-5.7}$~Gyr & 0.9~Gyr & 3.88 & 0.9~Gyr   \\
$\rm GV$ & $320^{+150}_{-100}$~Myr & 620Myr & 2.94 & 610~Myr   \\
$\rm BC$ & $1.0^{+0.3}_{-0.3}$~Gyr & 0.7~Gyr & 2.60 & 0.7~Gyr   \\
\hline
\end{tabular}

\begin{tabular}{p{17cm}}
$^{\rm a}$ The lifetimes of galaxy subsamples with $\rm{log(M_\ast/M_\odot)\ge10.0}$ estimated through the clustering results and a \citet{Tinker_2008} HMF. \\
$^{\rm b}$ The lifetimes of galaxy subsamples estimated through SED modeling by setting the key parameters (redshift, stellar mass, dust-corrected rest-frame UV colors and e-folding star-formation decline timescale $\tau$) to that of the observed subsamples, and leaving formation redshift $z_f$ as a free parameter. \\
$^{\rm c}$ The formation redshift of the subsample in SED modeling. \\
$^{\rm d}$ The peak of lifetime for each galaxy subsample in the Monte-Carlo simulation. \\
\end{tabular}

\label{tab:lft_model}
\end{table*}

\begin{figure}
\centering
\includegraphics[width=0.8\columnwidth, angle=90]{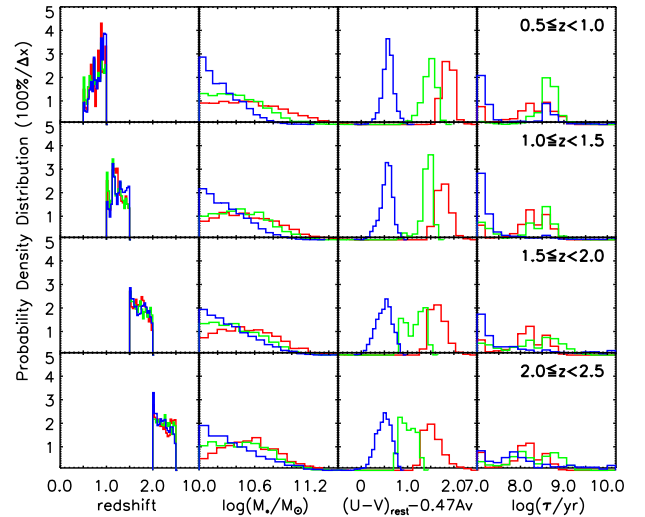}
\caption{The distributions of the redshift $z$, stellar mass $M_\ast$, rest-frame extinction-corrected UV color $\rm{UV_{rest,ext-cor}}$, and exponential star-formation decline timescale $\tau$ for galaxies in the COSMOS/UltraVISTA field. Random galaxies in the Monte-Carlo simulation are generated based on the same distributions of these galaxy parameters.}
\label{fig:hist_parameters}
\end{figure}

\begin{figure}
\centering
\includegraphics[width=0.8\columnwidth, angle=90]{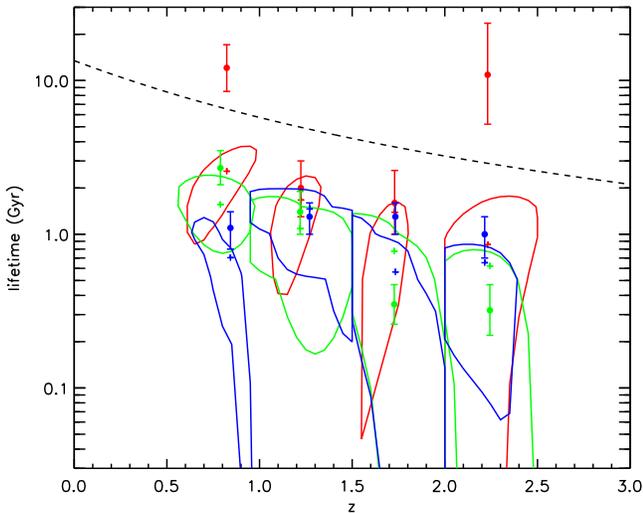}
\caption{The lifetimes estimated through clustering are compared with those from the SED modeling. The filled circles show the clustering-based lifetimes of the RS, BC, and GV galaxies with ${\log~(M_\ast/M_\odot)\ge10.0}$ in different redshift bins, which are the average lifetimes computed with the 6 different HMFs in Section~\ref{sect:dis_HMFs}. Each plus shows the lifetime estimated through the SED modeling by fixing the key parameters (i.e., redshift, stellar mass, extinction-corrected rest-frame UV color, and exponential star-formation decline timescale) to the median values of each respective observed galaxy subsamples. The color envelops show the $1\sigma$ lifetime distributions from the Monte-Carlo simulation for all galaxy subsamples, the peaks of which are very close to the pluses. The black dashed line shows the age of the universe as a function of redshift.
}
\label{fig:lft_model}
\end{figure}

To make our analyses more systematically, we also perform a Monte-Carlo simulation to show the lifetime distribution of the SED modeling in comparison with the clustering-based lifetime.
We randomly generate 10,000 data points in representation of the parameters of 10,000 random galaxies, ensuring that they have the same redshift ($z$), stellar mass (${M_\ast}$), rest-frame dust-corrected UV color ($\rm{UV_{rest,ext-cor}}$), and exponential star-formation declination timescale ($\tau$) distributions as each observed galaxy subsample selected with ${\log~(M_\ast/M_\odot)\ge10.0}$ in the COSMOS/UltraVISTA field.
The distributions of all these parameters for the random galaxy samples (same as the observed subsamples) are shown in Figure~\ref{fig:hist_parameters}.
For each random galaxy (having a set of four parameters, $z$, ${M_\ast}$, $\rm{UV_{rest,ext-cor}}$, $\tau$), we calculate its evolutional track in the color-mass diagram (Figure~\ref{fig:cmr_evo}) similarly to what we have done above for each galaxy subsample. The lifetime of each random galaxy is computed through the track. The peak of the simulated lifetime distribution is very close to the result of our SED modeling for each galaxy subsample, which is also tabulated in Table~\ref{tbl-5}. The $1\sigma$ distributions of the lifetimes of all random galaxies are plotted as the envelops surrounding their peaks in Figure~\ref{fig:lft_model}.

\subsection{Comparison of lifetime for different galaxy populations}\label{sect:lft_compare}

In general, the clustering-based lifetimes are very close to the results from the SED modeling as demonstrated in Figure~\ref{fig:lft_model}.
For the RS galaxies, lifetimes computed through the SED modeling are slightly smaller than that from clustering. This is because we define the lifetime of RS galaxies as the time interval since the galaxy crosses the GV region till the lower limit of the redshift bin of the respective observed subsample. But if we consider red sequence as a stable phase, the time that the galaxy stays in it can be much longer than expected if we do not consider other minor physical processes like rejuvenation.

RS galaxies generally have larger modeled lifetimes than BC and GV galaxies at all redshifts.
The SED modeled lifetimes of GV and BC galaxies are similar at $z>1$.
But at $z\le1$ the lifetime of GV galaxies grows rapidly to $\rm\sim1.6~Gyr$, while the lifetime of BC galaxies remains to be $\rm\sim700~Myr$.
At $z\ge1.5$, the clustering-based lifetimes of GV galaxies are shorter than BC galaxies at the same redshifts, but we do not observe this trend for their SED modeled lifetimes.
This may be due to the heavy degeneracy of stellar population parameters when constructing their color evolution.
Besides, the intrinsic uncertainties of the parameters also add to the difficulties in distinguishing the deviation of SED modeled lifetimes between the two galaxy populations.

If we consider the quenching scenario of star-forming galaxies, the lifetimes of BC/GV galaxies can be a rough estimate of their typical quenching timescales.
The clustering-based lifetimes in combination with the lifetimes estimated by the SED modeling have depicted an image of galaxy quenching for us.
In each redshift bin, RS galaxies are a stable already quenched population, and they have the longest lifetimes.
The distinct levels of lifetime of GV galaxies at $z\le1$ and $z>1$ may reflect that the quenching timescale of this transition galaxy population is different at higher and lower redshifts.
This is consistent with the result in \citet{Rowlands_2018}, where GV galaxies are found to follow a distinct quenching route at $z<1$ and $z\ge1$, which is defined as the ``fast and slow quenching route''.
At $z\ge1$, the GV galaxies quench very fast with a visible timescale of only a few hundred million years; while at $z<1$, the typical quenching timescale can be as long as several giga years.

As a comparison, the BC galaxies seem to follow a more constant quenching timescale independent of redshift, shown by their clustering-based lifetime in Figure~\ref{fig:lft_model}.
And it seems that the lifetime of BC galaxies modeled with a composite SED of exponentially declining star formation is slightly below their clustering-based lifetime at each redshift.
This is because we do not consider the newly born BC galaxies at each redshift bin when modeling the color evolution of BC galaxies using a single redshifted composite SED.
These newly born BC galaxies can delay the decrease of star formation of the entire BC population and extend their true lifetimes.

\subsection{What makes the lifetimes of GV galaxies short at $z\ge1.5$?}\label{sect:lft_GV_short}

At $z\ge1.5$, the shortest clustering-based lifetimes of GV galaxies among all galaxy populations may suggest that some additional physical processes are likely to join accelerating their quenching, since the SED modeling alone (only accounting for the effect of star formation) cannot clearly distinguish their shorter lifetimes from the larger lifetimes of BC galaxies.
One of the possibilities of accelerating quenching in GV galaxies is through the feedback of their central AGNs.
It has been found that AGN feedback has a great influence on regulating the star formation in galaxies \citep[e.g.,][]{Kormendy_2009,Feruglio_2010,Scholtz_2018}.
Since our subsamples have removed all the AGNs in the COSMOS/UltraVISTA catalog, it is diffcult to identify the remaining sources with potential AGN like features.

Nevertheless, some studies argue that morphological compaction may be necessary to trigger the central engine of AGN feedback \citep{Sales_2010,Dubois_2013,Diamond_2012}.
If this is the case, a fraction of the transition galaxies between the star-forming and quiescent populations should have compact structures, which is indeed discovered by a number of studies \citep[e.g.,][]{Barro_2013,Barro_2016,Kocevski_2017}.
Some recent relevant studies also found that quenching in the GV galaxies is likely to happen more rapidly in compact galaxies than the normal control group \citep[e.g.,][]{Powell_2017,NogueiraCavalcante_2019}.

We have checked the morphology of our galaxy subsamples.
We obtain the morphology information of our galaxy subsamples by cross-matching the COSMOS/UltraVISTA galaxies with the 3d-HST morphology catalog provided by \citet{Huertas-Company_2015}, who identified H-band morphologies of $\sim50,000$ galaxies in the five CANDELS fields through deep learning.
Then following the definition in \citet{Barro_2013}, we classify sources with $M_\ast/r_e^{1.5}\ge10.3$ as ``compact'' galaxies, where $r_e$ is the half-light effective radius of each galaxy.
This separation criterion was proven to be valid at $1.4<z<3.0$ in \citet{Barro_2013}.
Although the our subsamples extend to lower redshift bins, it can still be a rough estimate of the compact fraction of our subsamples.
We find that in all redshift bins, the fractions of compact galaxies are higher in GV galaxies than in BC galaxies.
Of the successfully matched galaxies, the fractions of GV galaxies with compact morphologies are 1.7\%, 15\%, 28\% at $z\sim0.8$, $z\sim1.2$ and $z\sim1.7$, which are $\sim3$ times the fractions of compact BC galaxies at these redshifts respectively.
At $z<1.5$, we may have under-estimated the compact fractions, since lower-redshift galaxies are less compact than they are at higher redshifts, and the morphology compactness classification criterion should also be correspondingly scaled down.
In the highest redshift bin of $2.0\le z\le2.5$, the discrepancy between the compact fractions of the two populations becomes smaller, with 19\% being compact GV galaxies and 13\% being compact BC galaxies.

We note that in some other studies AGN affecting star formation can happen in host galaxies of all morphological types \citep{Zubovas_2013}.
Some external processes such as major mergers \citep[e.g.,][]{Springel_2005a,Hopkins_2009,Kormendy_2013} or violent disk fragmentation \citep[e.g.,][]{Hopkins_2008,Martig_2009,Behrendt_2016,Bieri_2016} in star-forming galaxies can also trigger AGN feedback.
There are also studies about the infrared properties of galaxies of different structures, e.g., \citet{Chen_2020}, revealing that galaxies of different compactness share similar SFRs and dust temperatures, and that compact star-forming galaxies tending to be more quiescent than normal star-forming galaxies can turn out to be a ``selection effect''.
Thus, whether these galaxies had undergone a stage of morphological compaction in galactic scales before AGN feedback is still a topic under debate.

\subsection{Comparison between lifetime and typical gas depletion timescale for different galaxy populations}\label{sect:gas}

\begin{figure*}
\centering
\includegraphics[ width=0.75\columnwidth, angle=90]{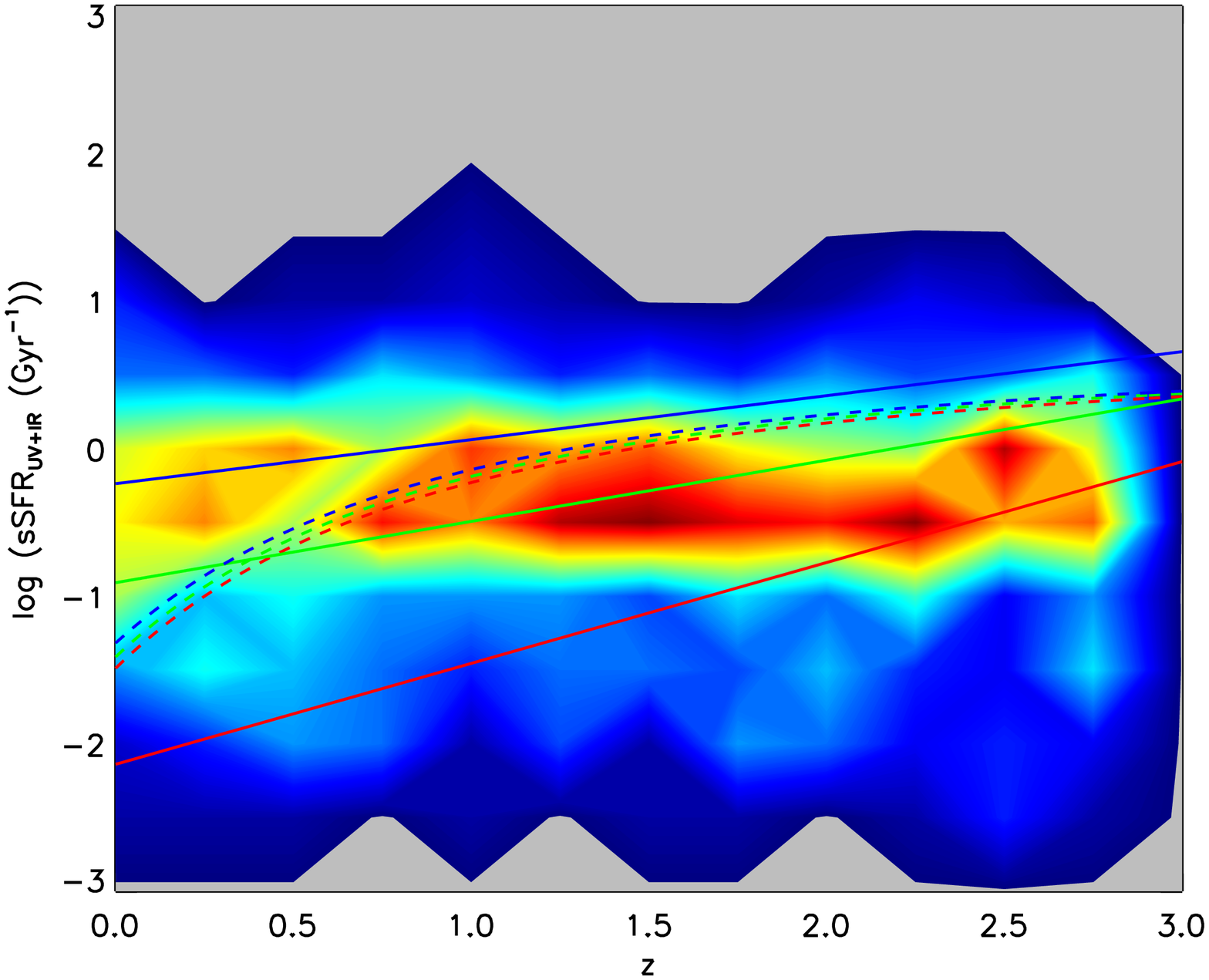}
\includegraphics[ width=0.75\columnwidth, angle=90]{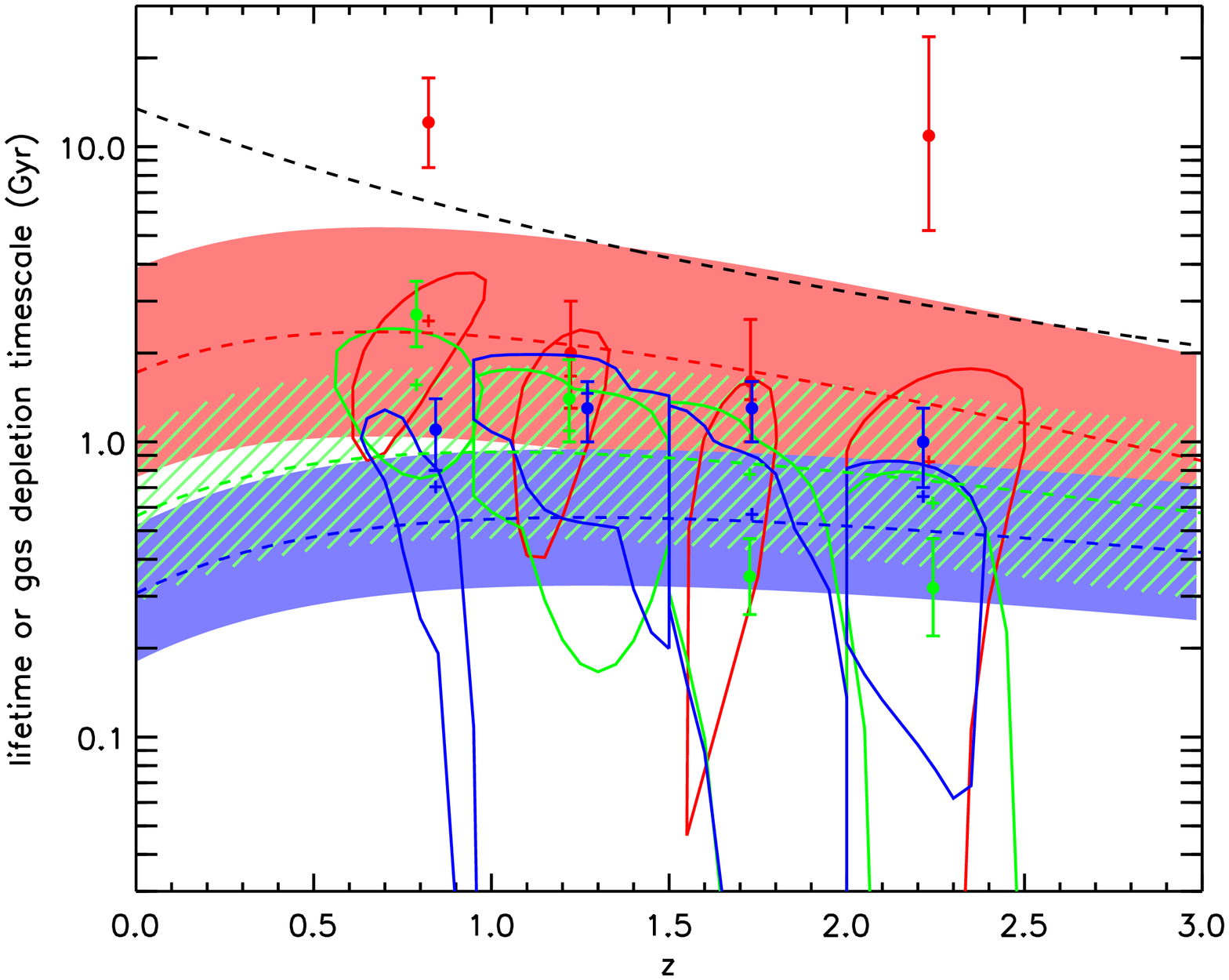}
\caption{ Left: the distribution of sSFR of COSMOS/UltraVISTA galaxies as a function of redshift.
Colors from red to blue depict the density gradient of galaxies distributed in the parameter plane.
The SFRs are calculated as a sum of contributions from the rest-frame UV and IR fluxes.
Dashed curves are the empirical sSFRs of the main-sequence galaxies derived using the prescription of \citet{Speagle_2014} (Equation \ref{Equ:MS_S14} in this paper) by fixing the stellar mass in the equation to the median $M_\ast$ of RS, GV and BC galaxies at all redshifts.
Solid lines show the power-law fitting of the sSFRs of our RS, GV and BC galaxies as a function of redshift.
The empirical $\rm sSFR-z$ relations of \citet{Speagle_2014} generally lie between the fitting results of our BC and GV galaxy subsamples at $0.5\le z\le3$.
Right: the comparison between the lifetimes of different galaxy subsamples selected with $\log(M/M_\ast)\ge10.0$ and the typical gas depletion timescales.
The red, green, and blue dashed curves show respectively the typical gas depletion timescales for the RS, GV, and BC galaxies derived using the script of \citet{Tacconi_2018}, with $\rm sSFR(MS,z,M_\ast)$ set to the empirical relation of \citet{Speagle_2014} (see Section~\ref{sect:gas} for details).
The colored shaded/line-filled regions show their $1\sigma$ ranges adopted as the maximums and minimums of $t_{\rm depl}$ when sSFRs and $M_\ast$ of the subsamples vary within $1\sigma$.
}
\label{fig:lft_gas}
\end{figure*}

We also compare the resulting lifetimes with the typical gas depletion timescales of the RS, GV and BC galaxies calculated with the script of \citet{Tacconi_2018}, who presented a fit for the galaxy gas depletion timescale, which is dependent on redshift, stellar mass, SFR, and effective radius.
The gas depletion timescale was derived as total molecular gas mass divided by SFR.
In their paper, they combined three methods in the measurement of molecular gas mass, i.e., the integrated CO line luminosity, FIR/submillimeter dust emission SED, and a single-band measurement of the dust emission at 1mm.
Then given an appropriate assumption of the conversion factor, the molecular gas mass can be obtained.
The SFRs in their paper were estimated based on FIR emission, MIR emission, and UV-to-NIR SED fitting in the order of decreasing preference.
We adopt their best error-weighted multi-parameter regression gas depletion timescale, which is dependent on redshift, SFR and stellar mass:
\begin{equation}\label{Equ:t_depl}
\begin{split}
{\leftline{$\log t_{\rm depl}(\rm Gyr)=0.002-0.37\times\log(1+z)-0.40\times$}}\\
{\rm\log (sSFR/sSFR(MS,z,M_\ast))+0.17\times(\log~M_\ast-10.7)},
\end{split}
\end{equation}
where $\rm sSFR(MS,z,M_\ast)$ is the specific SFR (sSFR) of the main-sequence galaxies as a function of redshift and stellar mass. Following \citet{Tacconi_2018}, we adopt the prescription of the star-formation main sequence proposed by \citet{Speagle_2014} in this study,
\begin{equation}\label{Equ:MS_S14}
\begin{split}
{\rm\leftline{$\log (sSFR(MS,z,M_\ast))=(-0.16-0.026t_c)\times$}}\\
(\log M_\ast+0.025)-(6.51-0.11t_c)+9, \\
\& \log t_c=1.143-1.026\log (1+z)-0.599\log^2 (1+z)\\
+0.528\log^3 (1+z).
\end{split}
\end{equation}
In this equation, $t_c$ is the cosmic time of a flat $\Lambda$CDM universe in units of Gyr.

Since the SFRs of COSMOS/UltraVISTA galaxies were derived with independent data in contrast to the statistical data used in \citet{Speagle_2014}, it is of necessity to check whether estimation of $\rm sSFR(MS,z,M_\ast)$ through Equation~\ref{Equ:MS_S14} is still valid for our galaxy subsamples.
In the left panel of Figure~\ref{fig:lft_gas}, we compare the $\rm sSFR(MS,z,M_\ast)$ of \citet{Speagle_2014} prescription with the sSFRs of our COSMOS/UltraVISTA galaxies.

For each RS, GV, or BC galaxy subsample, we calculate its median values of the stellar mass and SFR respectively. Then power-law fittings are provided to derive the stellar mass and SFR as a function of redshift for galaxies of a specific color (i.e., red, green or blue).
The fittings of sSFRs as a function of redshift for three galaxy populations are shown as solid lines with respective colors in the left panel of Figure~\ref{fig:lft_gas}.
The total SFR of each galaxy is determined via $\rm{SFR_{tot}=SFR_{UV,uncorr}+SFR_{IR}}$, which is presented in the \citet{Muzzin_2013a} catalog, where the ${\rm SFR_{IR}}$ is derived with the $\rm{24~\mu m}$ luminosity using the script of \citet{Kennicutt_1998}.
Dashed curves in the left panel of Figure~\ref{fig:lft_gas} show the $\rm sSFR(MS,z,M_\ast)$ of \citet{Speagle_2014} prescription by fixing the stellar mass in the equation to the median of our RS/GV/BC subsamples at each redshift.
It is clear that the $\rm sSFR(MS,z,M_\ast)$ values of \citet{Speagle_2014} prescription generally lie between the median sSFRs of our BC and GV galaxies at $0.5\le z\le3.0$.
Within the stellar mass range our subsamples covered, $M_\ast$ only has minor influence on the resulting $\rm sSFR(MS)~vs.~z$ relation for the \citet{Speagle_2014} prescription.
Thus we can expect the prescription of \citet{Speagle_2014} to be a good estimate of the sSFRs of the main-sequence galaxies in the COSMOS/UltraVISTA field, and is reliable in the computation of gas depletion timescales through Equation~\ref{Equ:t_depl}.

In the right panel of Figure~\ref{fig:lft_gas}, we compare the lifetimes of our galaxy subsamples with their typical gas depletion timescales.
The typical gas depletion timescales of the RS, GV, and BC galaxies as a function of redshift are shown as the color-coded dashed curves with their $1\sigma$ uncertainties shown in shaded/line-filled regions.
The boundary of the $1\sigma$ region of galaxies of a specific color is set as the maximum and minimum of the $t_{\rm depl}$ at each redshift when their sSFR and $M_\ast$ vary within $1\sigma$.

The gas depletion timescales of the RS and BC galaxies overlap with the regions where the lifetimes (no matter clustering based or SED modeled ones) of different populations lie in.
The stratification of gas depletion timescale with galaxy color is also in agreement with the separation of lifetimes between quiescent (red) and star-forming (blue/green) population.
It means that gas consumption by pure star formation is very possible to be the major cause of the transformation of galaxy color.

But we note that for the GV galaxies, the redshift evolution of the gas depletion timescale could not reflect the distinct behavior of their lifetime at higher and lower redshifts, as the gas depletion timescale of the GV galaxies computed through the script of \citet{Tacconi_2018} always lies between that of the RS and BC galaxies at whatever redshifts. This incongruity can be tuned if additional physical processes other than pure star formation would help accelerate the gas depletion of the GV galaxies, making their typical quenching timescale shorter than that of typical BC galaxies at high redshifts ($z>1.5$). One natural possibility is AGN feedback. The jets launched near the central supermassive black hole may quickly blow away the cold gas within the GV galaxies and stop the star formation in a rather short timescale \citep{Bower_2006,Cicone_2014,Schawinski_2014,Powell_2017,NogueiraCavalcante_2019}.

\subsection{Reproducing the redshift distribution of UV-selected galaxies}\label{sect:reproduce}

\begin{figure*}
\centering
\includegraphics[width=0.49\textwidth]{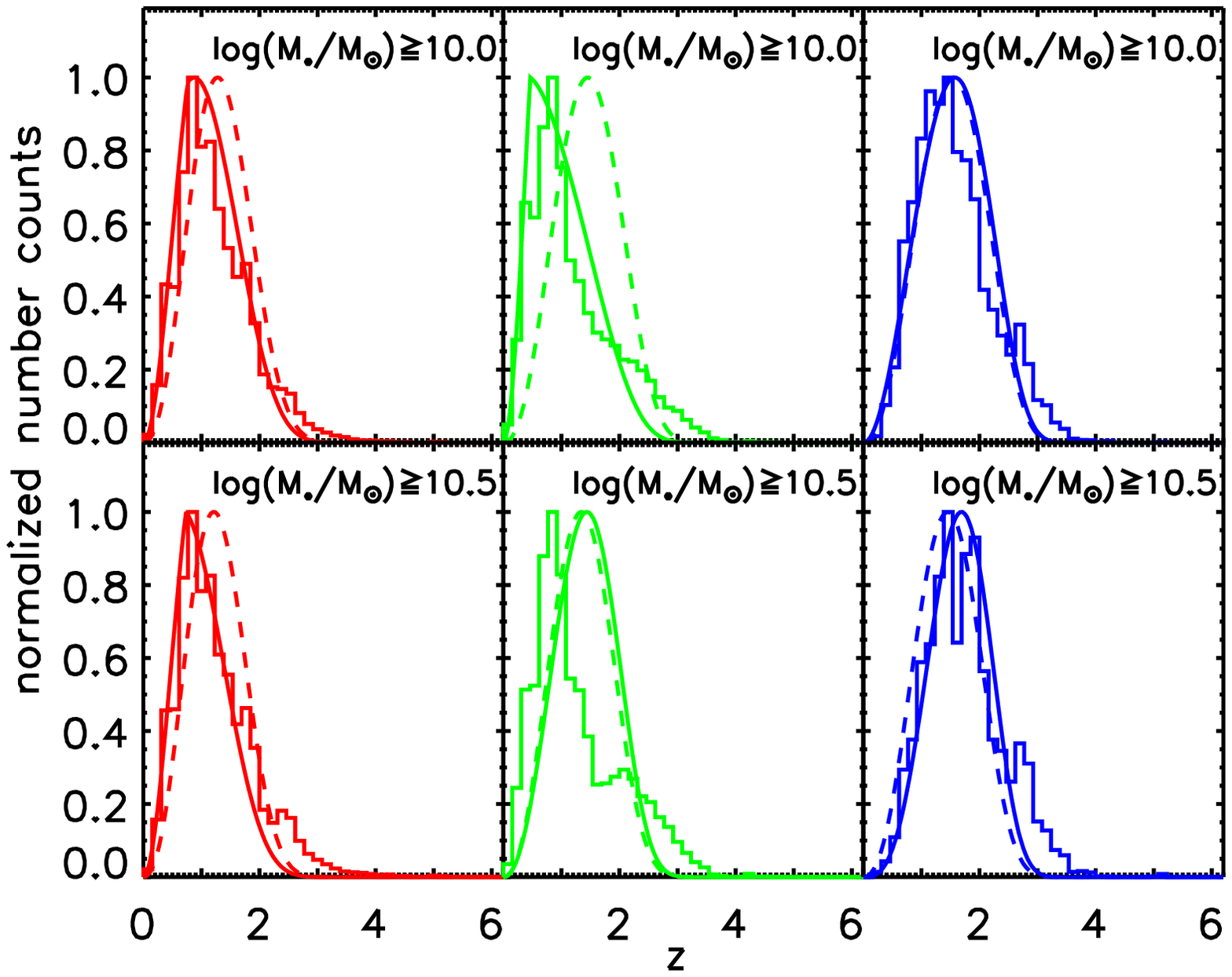}
\includegraphics[width=0.49\textwidth]{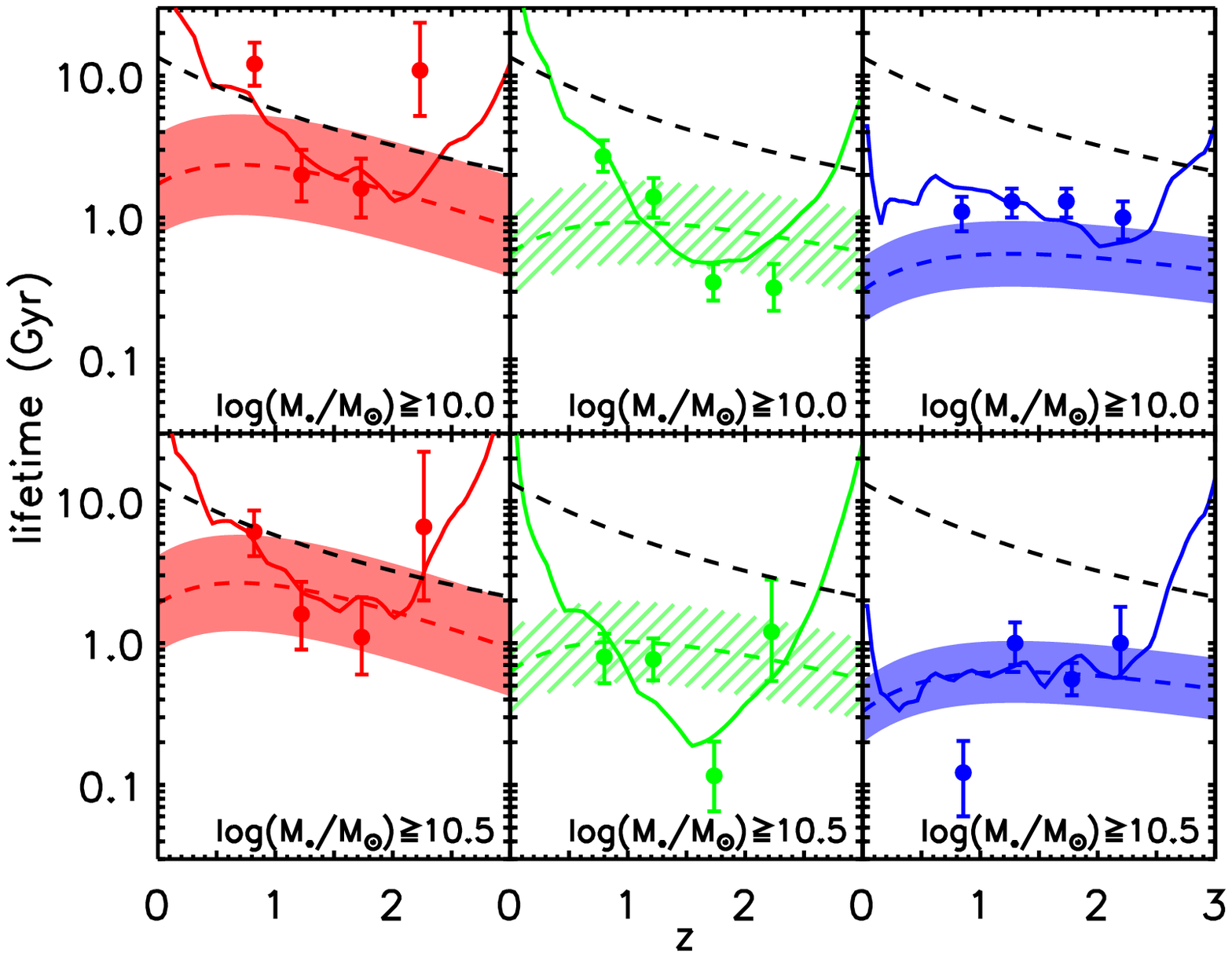}
\caption{Left: The redshift distributions of the RS, BC, and GV galaxies in the COSMOS/UltraVISTA field. The histograms show the redshift distributions of the observed RS, BC, and GV galaxies, whose redshifts are directly taken from the photometric-redshift catalog of \citet{Muzzin_2013a}. Upper panels show the redshift distributions of RS, GV and BC galaxies in the COSMOS/UltraVISTA field selected with $\log(M_\ast/M_\odot)\ge10.0$. Lower panels are the same, but for the galaxies selected with $\log(M_\ast/M_\odot)\ge10.5$. The curves in all panels show the estimated redshift distributions of the RS, BC, and GV galaxies through Equation~\ref{eq:Nz_thresh} assuming the \citet{Tinker_2008} HMF. For the solid curves, we use a power-law to fit the lifetimes of the RS, BC, and GV samples, respectively, to derive $t_{\rm sample}$; while for the dashed curves, we adopt the typical gas depletion timescales from \citet{Tacconi_2018} as $t_{\rm sample}$ (i.e., the color dashed curves in the right part of Figure~\ref{fig:lft_gas} and also shown in the right part of this figure). Right: The optimal parameters of $t_{\rm sample}$ in Equation~\ref{eq:Nz_thresh} to reproduce the redshift distributions of RS, BC and GV galaxies (colored solid curves), and their comparison with the clustering-based lifetimes (filled dots with error bars) and the typical gas depletion timescales from the script of \citet{Tacconi_2018} (colored dashed curves with $1\sigma$ uncertainties).}
\label{fig:pz_halo_galaxies}
\end{figure*}

Finally, we try to reproduce the photometric-redshift distributions of the RS, BC, and GV galaxies in the COSMOS/UltraVISTA field \citep{Muzzin_2013a} based on the HMF and the estimated lifetime of each subsample.
In the \citet{Hickox_2009} scenario, the activities of active galaxies and QSOs are related to the host DM halos. When the DM halo grows and reaches a certain mass ${M_{\rm halo}=M_{\rm thresh}}$, the activity of the host galaxy will be triggered (see Figure 16 of \citealt{Hickox_2009} for a schematic illustration of this picture). Thus, the differential number of halos crossing the mass threshold ${M_{\rm thresh}}$ as a function of redshift can be modeled as
\begin{equation}\label{eq:Nz_thresh}
\frac{dN_{\rm thresh}}{dz} \propto \frac{dn_{\rm halo}}{dM_{\rm h}}(M_{\rm thresh},z)\dot{M}_{\rm halo}(M_{\rm thresh},z)t_{\rm sample}\frac{dV}{dz},
\end{equation}
where $\frac{dn_{\rm halo}}{dM_{\rm h}}$ is the number density of DM halos per unit halo mass computed using the \citet{Tinker_2008} HMF, $\dot{M}_{\rm halo}$ is the typical growth rate of DM halos \citep{Fakhouri_2010}, $t_{\rm sample}$ is the lifetime of the RS, GV, or BC galaxies as a function of redshift and $\frac{dV}{dz}$ is the differential comoving volume over the survey area, respectively.
We derive $t_{\rm sample}$ by fitting a single power-law to the lifetimes of RS, GV, and BC galaxies computed with the \citet{Tinker_2008} HMF in four redshift bins, respectively.
Data points as well as parts of the fitting results exceeding the age of the universe at a given redshift (above the black dashed line in Figure~\ref{fig:lft_model}) will be scaled down to the age of the universe at the respective redshift.
If a specific phase (RS, GV, or BC) is triggered in a galaxy once the halo of the galaxy grows and reaches the halo mass of the respective phase at a given redshift, then the observed number density of galaxies in that phase will be expected to be proportional to $\frac{dN_{\rm thresh}}{dz}$. Using appropriate normalizations, we can reproduce the redshift distributions of the RS, BC, and GV galaxies and compare with the observations. \citet{Hickox_2012} also utilized this method to estimate the redshift distribution of sub-millimeter galaxies in their Section 4.5. We present our results in the left part of Figure~\ref{fig:pz_halo_galaxies}.

To make a comparison, we also substitute the lifetimes of the galaxy subsample $t_{\rm sample}$ by the typical gas depletion timescales of the RS, BC, and GV galaxies from \citet{Tacconi_2018} (the color dashed curves in Figure~\ref{fig:lft_model} and the right part of Figure~\ref{fig:pz_halo_galaxies}).
For the RS and BC galaxies, applying both the fitted lifetime and gas depletion timescale in Equation~\ref{eq:Nz_thresh} can reproduce their redshift distributions well. But for the GV galaxies, only applying the lifetime as $t_{\rm sample}$ can have a good reproduction of the redshift distributions. This is especially obvious in the stellar-mass bin of ${\log~(M_\ast/M_\odot)\ge10.0}$, where we have a larger amount of galaxies. The substitution of the lifetime of GV galaxies with their typical gas depletion timescale is very likely to overestimate the number of the GV galaxies at $z \sim 2$.

For a better visual identification of the difference between the clustering-based lifetimes and typical gas depletion timescales and to evaluate their performance on reproducing the redshift distributions for each galaxy population, we display the optimal parameters of $t_{\rm sample}$ (colored solid curves) in the right part of Figure~\ref{fig:pz_halo_galaxies}, corresponding to the parameters reproducing the exact redshift distributions (histograms in the left part of Figure~\ref{fig:pz_halo_galaxies}) of COSMOS/UltraVISTA galaxies in the \citet{Muzzin_2013a} catalog.
The reproduced redshift distributions for each galaxy population are properly normalized, such that the tracks of the parameters of $t_{\rm sample}$ can be scaled up/down appropriately to best-fit their measured clustering-based lifetimes.

It is clear that the optimal parameters reproducing the redshift histograms are very close to the clustering-based lifetimes for each galaxy population. Both of them show distinctive redshift evolution compared with the typical gas depletion timescales (dashed curves in the right part of Figure~\ref{fig:pz_halo_galaxies}).
For RS and BC galaxies, the evolutionary differences between the optimal parameters and the typical gas depletion timescales are insignificant over the redshift range we consider.
But for GV galaxies, the optimal parameters reproducing redshift histograms show great variations as redshift evolves, being well consistent with the measured clustering-based lifetimes, but completely different from the nearly constant gas depletion timescales computed through the script of \citet{Tacconi_2018}.

Interestingly, the optimal parameters reproducing the redshift distribution histogram of GV galaxies show minimums at the redshift of $z\sim1-2$, which is considered to be the peak of the global cosmic star formation and the peak of the global central black hole accretion \citep{Madau_2014,Heckman_2014,Delvecchio_2014,Sijacki_2015,Ma_2015,Zakamska_2016}.
This may be an implication that AGN feedback activities help accelerate quenching in these galaxies, making their lifetimes shorter than expected from the gas consumption by pure star formation.
The scenario that black hole feedback suppressing galaxy star formation at high masses is supported by a number of semi-analytic models and numerical hydrodynamic simulations \citep[e.g.,][]{Dubois_2014,Vogelsberger_2014,Somerville_2015a,Schaye_2015,Khandai_2015}.
Since most of these simulations are implemented with phenomenological models, further observations are needed to confirm this black hole negative feedback effect on the star formation of host galaxies.
Mini simulations on multi scales with precise physical assumptions could also help bridge the gap between stellar and cosmological scales \citep{McCarthy_2010,FaucherGiguere_2012,Torrey_2014,Lacey_2016,Weinberger_2017,Bower_2017}.
Observationally, debates about whether gas outflows driven by central black hole trigger or suppress galaxy star formation still exist \citep{Page_2012,Walch_2012,Silk_2013,Zubovas_2013,Maiolino_2017}.
Thus, it is currently difficult to make a firm conclusion that BH feedback would inevitably result in the suppression of galaxy star formation.

Nevertheless, if this scenario is indeed the case, the inconsistency between using clustering-based lifetime and typical gas depletion timescale to reproduce galaxy redshift distribution can be fully reconciled, since pure star formation is thus not the only way to consume the gas content within a GV galaxy at $z\sim2$. Other additional physical processes should also be considered. By effectively shortening the gas depletion timescales in these galaxies at $z\sim2$ through shocks or gas outflows driven by their central black holes, it is expected that the redshift distribution of the GV galaxies can be better reproduced.

\section{Summary}\label{sect:sum}

In this paper, we estimate the lifetimes of UV-selected galaxies in the COSMOS/UltraVISTA field at $0.5\le z\le2.5$. We first separate the RS, BC, and GV galaxies according to the color-separation criteria proposed by \citet{Wang_2017}. Then based on the clustering results presented in \citet{Lin_2019}, we estimate the lifetimes of these three galaxy populations at different redshifts.

We explore multiple factors that can influence the resultant lifetimes, including the choices of different HMFs, the width of redshift bins, the growth of DM halo, and the stellar mass of the sample. We conclude that applying different HMFs can slightly influence the lifetimes by $\rm{\approx0.2~dex}$, $\Delta z=0.5$ is large enough to estimate the lifetimes, and considering the growth of DM halo within each redshift bin has little effect on lifetime estimation. We also find that the lifetimes of all galaxy subsamples generally decrease with the increasing stellar mass, but this trend becomes weaker as the redshift increases.

The lifetimes of galaxy samples agree well with those from the SED modeling.
The RS galaxies have the longest lifetime compared to those of the BC and GV galaxies at all redshifts, indicating that the red sequence is likely a stable phase.
The BC galaxies have a nearly constant lifetime at different redshifts, which is around $\rm{1~Gyr}$. And the lifetime of the GV galaxies seems to be strongly dependent on redshift. At lower redshifts, the GV galaxies have longer lifetime than that of the BC galaxies, but this trend goes to the opposite when we focus on higher redshift bins, where the lifetime of the GV galaxies can be as short as several hundred million years. We also compare our result with the gas depletion timescale inferred from \citet{Tacconi_2018}. For the RS and BC galaxies, the lifetimes generally agree well with the respective gas depletion timescales, but for the GV galaxies, the gas depletion purely by star formation cannot fully explain the huge difference of their lifetimes at higher and lower redshifts. We conclude that the study on the lifetime reflects the quenching scenario of SFGs. At high redshifts, additional processes, such as AGN feedback, are required to accelerate the quenching of the GV galaxies.

Finally, we try to reproduce the redshift distributions of the UV-selected galaxies in the COSMOS/UltraVISTA field. Based on the \citet{Tinker_2008} HMF, the typical DM halo growth rate by \citet{Fakhouri_2010}, and the clustering-based lifetimes inferred in this paper, we can well reproduce the redshift distributions of the RS, BC, and GV galaxies. We also try to reproduce the redshift distributions of galaxies by substituting the clustering-based lifetimes with the SFR-dependent typical gas depletion timescales derived from \citet{Tacconi_2018}. Again, we find that we can well reproduce the redshift distributions of the RS and BC galaxies, but for the GV galaxies, the number density at $z\sim2$ seems to be overestimated. For GV galaxies, the optimal parameters reproducing redshift distributions are very close to the measured clustering-based lifetimes, but different from the nearly constant gas depletion timescales over redshifts. And we stress that, this disagreement can be reconciled if other additional physical processes rather than star formation, such as AGN feedback, aid depleting gas in the GV galaxies.

\section{acknowledgments}

We acknowledge the constructive work by Adam Muzzin presenting the catalog in the COSMOS/UltraVISTA field.
This work is supported by the National Natural Science Foundation of China (Nos. 11673004, 23002601, 1320101002, 11433005, 12025303, 11890693, and 11421303).
X.Z.L. and Y.Q.X. acknowledge support from the CAS Frontier Science Key Research Program (QYZDJ-SSW-SLH006) and the K.C. Wong Education Foundation.
G.W.F. acknowledges the support from Yunnan young and middle-aged academic and technical
leaders reserve talent program (No. 201905C160039), Yunnan ten thousand talent program-young top-notch talent and Yunnan Applied Basic Research Projects (2019FB007).

\bibliography{ms.bbl}

\end{document}